\documentclass[final,3p,times]{elsarticle}

\usepackage{amssymb}
\usepackage{graphicx,multirow,caption,subcaption,multicol,setspace,amsmath,wrapfig}
\usepackage{commath}
\usepackage[linesnumbered,ruled]{algorithm2e}
\usepackage{algorithmic}
\usepackage{adjustbox}
\usepackage{booktabs,caption}
\usepackage[flushleft]{threeparttable}
\usepackage{lineno}
\usepackage{enumitem}
\setlist{nosep}
\usepackage{rotating,caption}
\usepackage[table]{xcolor}
\usepackage{soul,xcolor}
\usepackage{amsfonts}
\usepackage{makecell,eqnarray}

\usepackage{tikz}
\usetikzlibrary{shapes,arrows,decorations.markings}
\usepackage[EULERGREEK]{sansmath}
\usepackage{floatrow}
\DeclareFloatFont{henry}{\fontfamily{ccr}\fontseries{m}}
\usepackage{float}
\floatstyle{plaintop}
\restylefloat{table}

\biboptions{numbers,sort&compress}

\DeclareMathOperator*{\argmin}{arg\,min}

\usetikzlibrary{shapes}
\usetikzlibrary{shapes.multipart}
\definecolor{CC1}{rgb}{0.09, 0.45, 0.27}
\definecolor{CC2}{rgb}{0.9,0.45,0.1}
\usetikzlibrary{chains,decorations.pathreplacing}
\usetikzlibrary{fit,positioning,shapes.geometric}
\usepackage{tikz}
\usetikzlibrary{shapes,arrows,decorations.markings}
\usetikzlibrary{arrows,calc}
\usetikzlibrary {arrows.meta} 
\usepackage{relsize}

\usepackage{nicematrix}
\NiceMatrixOptions{
code-for-first-row = \color{blue} ,
code-for-last-row = \color{blue} ,
code-for-first-col = \color{blue} ,
code-for-last-col = \color{blue}
}
\usepackage[]{sidecap}

\graphicspath{ {Images/} }

\makeatletter
\newcommand{\algorithmfootnote}[2][\footnotesize]{%
  \let\old@algocf@finish\@algocf@finish
  \def\@algocf@finish{\old@algocf@finish
    \leavevmode\rlap{\begin{minipage}{\linewidth}
    #1#2
    \end{minipage}}%
  }%
}

\journal{Neurocomputing}

\begin{document}
\begin{frontmatter}

\title{Co-evolution of Neural Architectures and Features\\ for Stock Market Forecasting: A Multi-objective Decision Perspective}

\author[label1]{Faizal Hafiz\corref{cor1}}
\address[label1]{SKEMA Business School, Université Côte d’Azur, Sophia Antipolis, France}
\ead{faizal.hafiz@skema.edu}
\cortext[cor1]{Corresponding author}

\author[label1]{Jan Broekaert}
\ead{jan.broekaert@skema.edu}

\author[label1]{Davide {La Torre}}
\ead{davide.latorre@skema.edu}

\author[label2]{Akshya Swain}
\address[label2]{Department of Electrical, Computer \& Software Engineering, The University of Auckland, New Zealand}
\ead{a.swain@auckland.ac.nz}

\begin{abstract}

In a multi-objective setting, a portfolio manager’s highly consequential decisions can benefit from assessing alternative forecasting models of stock index movement. The present investigation proposes a new approach to identify a set of non-dominated neural network models for further selection by the decision-maker. A new \emph{co-evolution} approach is proposed to simultaneously select the features and topology of neural networks (collectively referred to as \emph{neural architecture}), where the features are viewed from a topological perspective as input neurons. Further, the co-evolution is posed as a multi-criteria problem to evolve \emph{sparse} and \emph{efficacious} neural architectures. The well-known \emph{dominance} and \emph{decomposition} based multi-objective evolutionary algorithms are augmented with a \emph{non-geometric} crossover operator to diversify and balance the search for neural architectures across conflicting criteria. Moreover, the co-evolution is augmented to accommodate the data-based implications of distinct market behaviors prior to and during the ongoing COVID-19 pandemic. A detailed comparative evaluation is carried out with the conventional sequential approach of feature selection followed by neural topology design, as well as a \emph{scalarized} co-evolution approach. The results on the NASDAQ index in \emph{pre-} and \emph{peri-}COVID time windows convincingly demonstrate that the proposed co-evolution approach can evolve a set of non-dominated neural forecasting models with better generalization capabilities.

\end{abstract}


\end{frontmatter}

\section{Introduction}

In a complex multi-objective setting involving near-stochastic input data and disparate trade market behaviors stemming from the COVID-19 pandemic, a decision-maker (DM) looks to mitigate the effect of uncertain factors in portfolio decisions. Structured information and oversight from forecasting models are essential supporting tools in such circumstances. This study, therefore, aims to obtain a set of Pareto-optimal forecasting models for stock index movement, which balance multiple criteria of model complexity with prediction performance over distinct market behaviors separated by the outbreak of the COVID-19 pandemic. 

This investigation, in particular, focuses on the design of neural forecasting models with input features derived from the \emph{technical analysis}, which essentially aggregate the information contained in the historical time-series of basic trading data (\textit{e.g.}, \textit{daily open} and \textit{close} index values). The motivation for this research direction is two-fold: (1) technical analysis is well-suited for short-term forecasting as it depends only on regularly available basic trading information and is arguably the most common among the existing approaches~\cite{Bustos:2020,Kumbure:Lohrmann:2022,Htun:Biehl:2023}. (2) Artificial Neural Networks (ANN) remained a benchmark technique since the early adoption of machine learning models in stock market forecasting~\cite{Henrique_EtAl_Review_Market_prediction_snooping_2019,Kumbure:Lohrmann:2022}. Further, while the flexible topology of ANN can capture possibly non-linear information in stock forecasting data, several issues related to efficient network design, such as optimal selection of topology and input features, still warrant attention, \textit{e.g.}, see~\cite{Atsalakis:Valavanis:2009,Hussain:Knowles:2008,Lam:2004,Versace:Bhatt:2004}, which demonstrate that such selection is likely to improve forecasting performance.

The forecasting performance of any neural model is critically dependent on its \emph{topological} design, which encompasses \emph{depth} (number of hidden layers), \emph{size} (number of neurons in each hidden layer), and the choice of the \emph{activation function}. The optimal topology for a given application and dataset requires an adjusted network complexity that balances the \emph{bias-variance} trade-off~\cite{Yao:1999,Wang:Xu:2018}, \textit{e.g.}, an overly \textit{sparse} network may lead to high bias errors, whereas an \emph{over-complex} network may over-fit and thus lead to increased generalization errors. Therefore, an empirical topological design based on a \textit{trial-and-error} or a \textit{rule-of-thumb}, which is often employed in existing stock market forecasting models (see~Section~\ref{subsec:nas} for details), is likely to yield a suboptimal performance. The other crucial design factor is \emph{feature selection} which aims to identify and remove \emph{irrelevant} and \emph{redundant} input features (~see~\cite{Kohavi:John:1997} for the details on \emph{feature relevance}), to reduce the input dimensionality, and often to improve the forecasting performance~\cite{Kohavi:John:1997,Guyon:Isabelle:2003,Hafiz:Swain:2018}, as many technical indicators tend to provide overlapping information and may become redundant~\cite{Peng:Albuquerque:2021,Htun:Biehl:2023}. While recent stock forecasting literature focuses on feature selection (see Section~\ref{subsec:featsel}), most depend either on feature selection \textit{filters}~\cite{Kohavi:John:1997,Guyon:Isabelle:2003} or on dimensionality reduction through principle component analysis, which often neglects non-linear interactions among features. To bridge this gap and to simultaneously address the selection of features and topology, this study pursues a new topological perspective where the selection of features is viewed, \emph{ab initio}, as the selection of \emph{input} neurons. Such integration can yield better results for both feature and topology selection as follows: the reduction in input dimensionality associated with feature selection encourages the exploration of relatively parsimonious topologies. Similarly, topological selection aids the search for feature subsets by providing direct access to classification performance; such direct estimates of feature subset efficacy are often more accurate than indirect statistical estimates; see \emph{feature selection filters} and \emph{wrappers} in~\cite{Kohavi:John:1997,Guyon:Isabelle:2003,Hafiz:Swain:2018}. It is worth emphasizing that, despite these advantages, most forecasting models focus on a conventional disjoint or \emph{sequential} neural design, where feature selection is followed by an empirical selection of neural topology~\cite{Asadi2012,Qiu:Song:2016,Zhong:Enke:2017,Tsai:Hsiao:2010,Peng:Albuquerque:2021}, which is likely to yield sub-optimal models.

We approach \emph{neural architecture} as a combination of a \emph{feature subset} and a \emph{neural topology} based on the aforementioned topological perspective. This allows us to formulate the multi-objective \emph{co-evolution} problem, which aims to identify the Pareto optimal set of \emph{efficacious} and \emph{parsimonious} neural architectures. Further, it is easy to follow that the search space for the multi-objective \emph{co-evolution} problem contains all the possible combinations of feature subsets as well as neural topologies (discussed later in Section~\ref{sec:Multi-objective_and_Co-evolution_ANN_search}). This search space is known to be multi-modal, deceptive, and noisy~\cite{Yao:1999,Guyon:Isabelle:2003}. To this end, a combination of Multi-objective Evolutionary Algorithms (MOEAs) and an \emph{a posteriori} decision making tool is proposed as the overall search framework. In particular, MOEA identifies a set of non-dominated neural architectures first, which represent a different degree of trade-off over parsimony and forecasting performance. Next, a combination of multiplicative preference relations~\cite{Zhang:Chen:2004} and a multi-criteria tournament~\cite{Parreiras:Vasconcelos:2009} is proposed as the \textit{a posteriori} decision support tool to select a particular neural architecture as per the preferences of the Decision Maker (DM).

To summarize, with the motivations for improved neural forecasting models for stock market movement, the core contributions of this investigation are as follows:
\begin{itemize}[leftmargin=*]
\item Simultaneous optimization of features and neural topology is proposed under the \emph{co-evolution} framework. The architectural complexity/parsimony is pursued under a multi-objective setting to evolve parsimonious 
neural architectures with better generalization capabilities.

\item Forecasting of the NASDAQ index is being considered during the COVID-19 pandemic. It is shown that the optimal architectural design for disparate market behaviors prior to and within the pandemic can be inconsistent to some degree and is addressed by balancing forecasting performances over \textit{pre-} and \textit{within-COVID} periods.

\item A search framework consisting of MOEA and \emph{a posteriori} decision support tool is proposed to identify neural architectures under the proposed co-evolution environment. The impact of search algorithms is investigated by considering two well-known MOEAs based on distinct search philosophies: \emph{dominance} based NSGA-II~\cite{Deb:Pratap:2002} and hybrid \emph{decomposition-dominance} based EAGD~\cite{Cai:Li:2015}. Further, NSGA-II is augmented by introducing a \emph{non-geometric} crossover operator~\cite{Ishibuchi:Tsukamoto:2010} to encourage diversity of the identified neural architectures. 
\end{itemize}

The efficacy of the proposed co-evolution is demonstrated by considering a total of $21$ different neural architecture design approaches, which are broadly categorized into three comparative \emph{neural design baselines} (see Section~\ref{sec:compeval}). The results of the comparative evaluation demonstrate a statistically significant improvement in the forecasting performance with the proposed co-evolution approach.
 
To develop our forecasting model and arguments, the rest of this article is organized as follows: In Section~\ref{sec:literature_review}, the related developments in the literature are screened. Section~\ref{sec:Model_outline} outlines the forecasting model, classification performance metrics, and COVID-19 related data segmentation. This is followed by the formulation of the proposed \emph{co-evolution} problem in~Section~\ref{sec:Multi-objective_and_Co-evolution_ANN_search}. The search framework consisting of Multi-objective Evolutionary Algorithms (MOEAs) and the  \textit{a posteriori} decision support tool is discussed next in Section~\ref{sec:Multi-objective_search}. Section~\ref{sec:compeval} details three distinct neural design baselines, which are considered for comparative evaluation purposes. The results of this investigation are discussed in Section~\ref{sec:results}. Finally, Section~\ref{sec:conclusions} provides a brief discussion and conclusions about the proposed \emph{co-evolution} approach and the corresponding search framework.

\section{Background and Related Works}
\label{sec:literature_review}

\subsection{Feature Selection}
\label{subsec:featsel}

Feature selection is one of the fundamental problems of machine learning, and it involves a \emph{sparse} selection of \emph{relevant} features from the given set of input features~\cite{Guyon:Isabelle:2003,Kohavi:John:1997,Hafiz:Swain:2018}. Most feature selection approaches can be categorized into either \emph{filters} or \emph{wrappers}. This distinction arises mainly from the estimation used to evaluate the classification performance; filters typically rely on indirect statistical or information theory based estimates, whereas wrappers depend directly on the performance of an underlying classifier. While wrappers are computationally expensive, they tend to be more accurate. We refer to~\cite{Guyon:Isabelle:2003,Kohavi:John:1997} for a detailed discussion on feature \emph{relevance}, \emph{redundancy} as well as filters and wrappers.

Most of the existing stock forecasting models rely on \emph{filters} to indirectly estimate feature relevance and/or redundancy, which include but are not limited to correlation criteria~\cite{Li:Zhang:2022,kumar:Meghwani:2016,Zbikowski:2015,Oliveira:Nobre:2013,Asadi2012,Huang2009}, mutual information~\cite{Sun:Xiao:2019} and information gain~\cite{Enke:2004}. In comparison, feature selection wrappers have received relatively less attention~\cite{Lee:2009,Weng:Ahmed:2017,Peng:Albuquerque:2021,Inthachot:Boonjing:2016,Liu:Wang:2019}. Peng \textit{et al.}~\cite{Peng:Albuquerque:2021} considered two wrappers, sequential forward floating search and tournament selection, with logistic regression as the underlying classifier for a day ahead movement prediction of seven stock indices. Lee~\cite{Lee:2009} proposed a hybrid filter-wrapper feature selection for a day-ahead movement prediction of the NASDAQ index. In particular, an F-score based filter is used first to prune the feature set, which is followed by a greedy sequential forward search (wrapper) with SVM as the underlying classifier. In~\cite{Weng:Ahmed:2017}, a combination of recursive feature elimination wrapper and SVM is used to reduce a full feature set derived from technical analysis as well as other exogenous sources. In~\cite{Inthachot:Boonjing:2016,Liu:Wang:2019}, a genetic algorithm based wrapper with ANN as the underlying classifier was proposed. Another popular approach among the existing forecasting models is feature extraction using Principal Component Analysis (PCA)~\cite{Zhong:Enke:2017,Zhong:Enke:2017b,Persio:Honchar:2016,Tsai:Hsiao:2010}, where the dimensionality reduction is achieved by retaining a few linear combinations of features (principals) that account for maximal data variance. The transformation of data from a higher to a lower dimensional space is, however, associated with a loss of \textit{interpretability} and may not be desirable. Further, different ensembles of feature selection techniques have also been explored~\cite{UlHaq2021,Alsubaie:Hindi:2019,Tsai:Hsiao:2010}. The objective of such ensembles is to complement individual feature selection techniques. For instance, Tsai and Hsiao~\cite{Tsai:Hsiao:2010} derived different ensembles through the union and intersection of reduced subsets identified using genetic algorithm (wrapper), information entropy (decision trees) and PCA. 

To summarize, while feature selection is receiving increasingly more attention in stock market forecasting, the focus is primarily either on filters or PCA, which tend to neglect nonlinear feature interactions. Further, feature selection is mostly approached as a uni-objective problem focusing primarily on classification performance. However, it is essentially a multi-objective problem as its two key objectives, \emph{subset sparsity} and \emph{improved classification performance}, are at least partially conflicting. Consequently, an optimal feature subset which minimizes both objectives simultaneously may not exist; instead, there may exist a set of non-dominated feature subsets which represent a different degree of trade-off among these objectives. 

\subsection{Neural Architecture Search}
\label{subsec:nas}

Before we review the existing neural architecture designs, it is pertinent to discuss the implications of neural network depth briefly. While shallow and deep neural architectures have been pursued to forecast stock index movement~\cite{Bustos:2020,Kumbure:Lohrmann:2022,Sezer:Gudelek:2020,Rundo:Trenta:2019,Henrique_EtAl_Review_Market_prediction_snooping_2019}, the selection of the optimal architectural complexity depends on the information in the data~\cite{Yao:1999,he2015deep}. When a network architecture is excessively \emph{deep} with respect to the statistical information in the data, the initial few network layers will optimally capture the relation while the remaining layers simulate the \emph{identity} function~\cite{he2015deep}. Following this perspective and accommodating the fact that the dynamic information contained within technical indicators is often considered \emph{weak} owing to the \emph{adaptive market} hypothesis \cite{Fama1965,Bustos:2020,Kumbure:Lohrmann:2022,Henrique_EtAl_Review_Market_prediction_snooping_2019}, this study pursues shallow neural architectures with a few hidden layers for stock index forecasting. We refer to~\cite{Bustos:2020,Kumbure:Lohrmann:2022,Sezer:Gudelek:2020,Rundo:Trenta:2019} for a detailed treatment of deep neural network-based forecasting.

The neural architecture can be approached as a set of design choices for the hidden layers, \textit{e.g.}, the number of neurons (size) and layers (depth). The selection of an appropriate topology has been the focus of active research since the early stages of neural network development~\cite{Yao:1999,Stathakis:2009}. Over the years, several \emph{rules-of-thumb} have been proposed which determine the size of hidden layers as a function of the number of inputs/features, outputs and training samples~\cite{Stathakis:2009,Hafiz:Broekaert:2021}, \textit{see}~Section~\ref{sec:baseline1}. Network pruning is another common approach wherein the topological complexity is gradually reduced by identifying and pruning redundant neurons, \textit{e.g.} see~\cite{Yao:1999}. The selection of the optimal topology is, however, often application-dependent, and an empirical design following a rule-of-thumb or trial-and-error is likely to yield suboptimal results~\cite{Yao:1999,Stathakis:2009,Hafiz:Broekaert:2021}.

It is interesting to note that most stock market forecasting models based on neural networks rely on empirical approaches for topological design~\cite{Peng:Albuquerque:2021,UlHaq2021,Liu:Wang:2019,Zhong:Enke:2017b,Zhong:Enke:2017,Inthachot:Boonjing:2016,Asadi2012,Tsai:Hsiao:2010,Lam:2004,Enke:2004,Kim2003,Cao:Tay:2003}. Typically, the size of the hidden layer is adjusted by limited \emph{trial-and-error} and cross-validation~\cite{Peng:Albuquerque:2021,UlHaq2021,Zhong:Enke:2017b,Zhong:Enke:2017,Asadi2012,Tsai:Hsiao:2010,Lam:2004,Enke:2004,Kim2003,Cao:Tay:2003}. The other empirical approach uses rules of thumb to determine the number and size of hidden layers~\cite{kumar:Meghwani:2016,Oliveira:Nobre:2013,Lee:2009,Olson:Mossman:2003}. A few existing approaches focus on meta-heuristics, such as genetic algorithm, to determine optimal network weights~\cite{Hu:tang:2018,Qiu:Song:2016,Chang2012}; however, the topology is still selected empirically. In~\cite{Lei2018}, a rule of thumb determines the initial topological design for a wavelet neural network, which is subsequently pruned using a rough set-based approach. In comparison, the optimization of neural topology has received relatively less attention, \textit{e.g.}, see~\cite{Farahani:2021,Gocken:2016,Kim:Shin:2007,Versace:Bhatt:2004}. Versace \textit{et al.}~\cite{Versace:Bhatt:2004} focused on optimizing the number of components in PCA along with the size of the hidden layer using genetic algorithm (GA). In~\cite{Gocken:2016,Farahani:2021}, GA was used to select features as well as the size of hidden the layer for a single hidden layer neural network. Similarly, Kim and Shin~\cite{Kim:Shin:2007} focused on optimizing input delays and the hidden layer size of time-delayed neural networks. 

The selection of neural topology is still under-explored in the stock forecasting models; while a few investigations in~\cite{Farahani:2021,Gocken:2016,Kim:Shin:2007,Versace:Bhatt:2004} focus on selecting a part of neural architecture, such as features and hidden layer size, the other aspects, such as the number of hidden layers and selection of activation function, have not been considered. Further, the complexity of neural architectures is neither explicitly defined nor pursued in any of these investigations. Our earlier investigation in~\cite{Hafiz:Broekaert:2021} demonstrated that such explicit formulation of neural complexity is key to balancing the parsimony and efficacy of the neural architectures.

\section{Forecasting Procedure}
\label{sec:Model_outline}

\subsection{Neural Forecasting using Technical Indicators}
\label{subsec:NeuralNet_Forecasting}

This study uses technical indicators to forecast a day-ahead stock index movement. Technical indicators are designed as functions of fundamental daily trading quantities (\textit{i.e.}, trading volume, open, close, intra-day high and low) to identify trends or turning points in historical index data, which would subsequently support trading decisions~\cite{Huang2009, kumar:Meghwani:2016,Bustos:2020,Peng:Albuquerque:2021,Hafiz:Broekaert:2021}. In particular, the neural forecasting model aims to capture relations between technical indicators and the movement of the index. In particular, a total of 24 distinct technical indicators shown in Table~\ref{table:technical_indicators} are considered in this study to capture various trends and turning points in the historical index data. These indicators have been selected on the basis of earlier investigations in~\cite{Huang2009, kumar:Meghwani:2016,Bustos:2020,Peng:Albuquerque:2021,Hafiz:Broekaert:2021,Kumbure:Lohrmann:2022}. The selected indicators include both \emph{trend indicators} and \emph{oscillators}. The trend indicators like Moving Average and Momentum (see Table~\ref{table:technical_indicators}) have been developed to identify the direction of movement. In contrast, the oscillators have been developed to recognize turning points by identifying over-bought and over-sold, \textit{e.g.}, Relative Strength Index and William's oscillator in Table~\ref{table:technical_indicators}. We refer to~\cite{Bustos:2020} for a detailed discussion on technical indicators. It is worth noting that each indicator essentially summarizes the index behavior over a period of the past few days, which is denoted by $\tau$ in Table~\ref{table:technical_indicators}. While the maximum value of $\tau$ is typically limited to 30 days, the exact value of $\tau$ is often set empirically \cite{Huang2009, kumar:Meghwani:2016,Bustos:2020,Peng:Albuquerque:2021}. Given that the selection of $\tau$ is likely to affect forecasting performance (see~\cite{Shynkevich:McGinnity:2017}), and there is no consensus on its optimum value, most of the indicators in this study are determined over different values of $\tau$, as shown in Table~\ref{table:technical_indicators}. Hence, multiple \emph{features} are obtained from most technical indicators. 

The features extracted from technical indicators are subsequently used as the \textit{inputs} to a \textit{feed-forward neural network} (see Section~\ref{sec:Multi-objective_and_Co-evolution_ANN_search}). The neural network is trained via supervised learning to predict \emph{a day-ahead index movement}, $y(t+1)$, which is encoded as a binary classifier:
\begin{equation}
    \label{eq:Ypred}
    y(t+1) = \begin{cases} 
            1, \quad \textit{if} \quad C(t+1)-C(t)>0,\\
            0, \quad \textit{otherwise}
    \end{cases}
\end{equation}
\begin{table*}[!t]
  \centering
  \small
  \caption{Technical Indicators$^{\dagger\dagger}$}
  \label{table:technical_indicators}
  \begin{adjustbox}{max width=0.92\textwidth}
  \begin{threeparttable}
    \begin{tabular}{lll}
       \toprule
       \textbf{Financial Indicator} & \textbf{Parameters}  & \textbf{Expression}  \\[0.5ex] 
       \midrule
       \makecell[l]{Opening, Highest Intra-day,\\Lowest Intra-day, \& Closing Price}   & -- &  $[O(t), H(t), L(t),  C(t)]$ \\[0.5ex]
     
       Moving Average & $\tau=[5,10,15,20]$ &  $\mathit{MA}_\tau (t) = \displaystyle \sum \limits_{j=t-\tau+1}^t \frac{C(j)}{\tau}$ \\[2ex]
        
       Exponential Moving Average & $\tau=[5,10,15,20]$, $\alpha = \frac{2}{\tau+1} $ &  $\mathit{EMA}_\tau (t) =  \alpha  C(t) +(1-\alpha) \mathit{EMA}_{\tau} (t-1),  \ \ \   \mathit{EMA}_\tau (1) = C(1)   $ \\[0.5ex]
        
        Relative Strength Index & $\tau=[5,10,15,20]$ & $\mathit{RSI}_\tau(t) = \displaystyle 100 \times \frac{\mathit{UPC}_\tau(t)/\mathit{UD}_\tau(t)}{\mathit{UPC}_\tau(t)/\mathit{UD}_\tau(t)+ \mathit{DPC}_\tau(t)/\mathit{DD}_\tau(t)}$\\[2.5ex]
        
        Stochastic Index, K & $\tau=[5,9]$ & $  K_\tau(t) = \displaystyle \frac{2}{3} K_\tau(t-1)+\frac{1}{3} \frac{C(t) -\mathit{HH}_\tau(t) }{\mathit{HH}_\tau (t) -\mathit{LL}_\tau (t)} $ \\[2.5ex]
        
        Stochastic Index, D & $\tau=[5,9]$ & $ D_\tau(t) = \frac{2}{3} D_\tau(t-1)+\frac{1}{3}K_\tau (t) $ \\[0.5ex]
        
        \makecell[l]{Moving Average\\Convergence-Divergence}  & $\tau=9$, $\alpha = \frac{2}{\tau+1} $ &  $\mathit{MACD}_\tau(t) =  (1-\alpha) \mathit{MACD}_\tau(t-1)+\alpha(\mathit{EMA}_{12} (t) - \mathit{EMA}_{26}(t))$\\[1ex]
        
        Larry Williams’ Oscillator & $\tau=[5,10,15,20]$ & $ \mathit{WR}_\tau(t) = 100 \times \displaystyle \frac{\mathit{HH}_\tau(t) - C(t) }{\mathit{HH}_\tau (t) -\mathit{LL}_\tau (t)}$\\[2ex]
        
        Psychological Line & $\tau=[5,10,15,20]$ & $\mathit{PSY}_\tau(t) =  100 \times \displaystyle\frac{\mathit{UD}_\tau(t)}{\tau}$\\ [2ex]
        
        Price Oscillator  &   $x = [5,10,15]$, $y=[10,15,20]$    &   $\mathit{OSCP}_{x,y}(t) = \displaystyle \frac{\mathit{MA}_x(t) - \mathit{MA}_y(t)}{\mathit{MA}_x(t)}$   \\ [2ex]
        
        $^\dagger$Directional Indicator Up & $\tau=[5,10,15,20]$  & $\displaystyle \mathit{+DIS}_\tau(t) = \Big(\mathit{+DM_{\tau}}(t)/ \mathit{TRS_{\tau}}(t) \Big) \times 100$\\ [1.5ex]
        
         $^\dagger$Directional Indicator Down & $\tau=[5,10,15,20]$  & $\displaystyle \mathit{-DIS}_\tau(t) = \Big(\mathit{-DM_{\tau}}(t)/ \mathit{TRS_{\tau}}(t)\Big) \times 100 $ \\[1.5ex]
        
        Bias & $\tau=[5,10,15,20]$ &  $\mathit{BIAS}_\tau(t) = 100 \times \displaystyle \frac{C(t)-\mathit{MA}_\tau(t)}{\mathit{MA}_\tau(t)}$\\[2ex]
        
        Volume Ratio & $\tau=10$ & $\mathit{VR}_\tau(t)  = \mathit{UV}_\tau(t)/\big(\mathit{UV}_\tau(t) + \mathit{DV}_\tau(t)\big)$  \\[1ex]
        
        A ratio & $\tau=20$ & $  AR_\tau(t) =    \sum_{j=t-(\tau-1)}^t \ H(j)-O(j) \Big/ \sum_{j={t-(\tau-1)}}^t \ O(j)-L(j)$  \\[1.5ex]
        
        B ratio  & $\tau=20$ & $BR_\tau(t) =    \sum_{j=t-(\tau-1)}^t / H(j)-C(j) \Big/ \sum_{j={t-(\tau-1)}}^t C(j)-L(j)$   \\[1.5ex]
        Lowest Low          & $\tau = 10$ &  $\mathit{LL}_\tau(t)  = \min\big\{L(t-\tau), \cdots, L(t-1) \big\} $ \\  [1.5ex]
        Highest High        & $\tau = 10$ &  $\mathit{HH}_\tau(t)  = \max\big\{H(t-\tau), \cdots, H(t-1)\big\}$ \\  [1.5ex]
        Median Price        & $\tau = 10$ &  $\mathit{MP}_\tau(t)  = {\rm med}\big\{C(t-\tau), \cdots, C(t-1)\big\} $ \\  [1.5ex]
        Average True Range  & $\tau = 10$ &  $\mathit{ATR}_\tau(t) =  \Big(\mathit{ATR}_\tau(t)\cdot (\tau-1) + \mathit{TR}(t)\Big)/\tau$  \\  [1ex]
        \makecell[l]{Relative Difference\\ in Percentage} &  $\tau=[5,10,15,20]$ & $\mathit{RDP}_\tau(t)= 100 \times \displaystyle\frac{C(t)-C(t-\tau)}{C(t-\tau)}$  \\  [2ex]
        Momentum            &   $\tau=[5,10,15,20]$ & $\mathit{MTM}_\tau(t) = C(t) - C(t-\tau)$  \\  [1.5ex]
        Price Rate of Change &   $\tau=[5,10,15,20]$ & $\mathit{ROC}_\tau(t) = \displaystyle 100 \times \frac{C(t) }{C(t-\tau)}$  \\  [2.5ex]
        $^\ddagger$Ultimate Oscillator &   $(x,y,z) = [10, 20, 30]$ & $\mathit{UO}_{x,y,z}(t) = \displaystyle \frac{100}{4+2+1} \Big(4\, \mathit{AVG}(x)+ 2\, \mathit{AVG}(y) +   \mathit{AVG}(z) \Big) $  \\  [2.5ex]
        Ulcer Index     & $\tau = 14$ &  $\mathit{Ulcer}_\tau(t) = \sqrt{\sum_{k=1}^\tau R_k(t)^2/\tau } $, $ R_k(t) = \displaystyle \frac{100}{\mathit{HH}(t-k)} \big(C(t) - \mathit{HH}(t-k)\big)$\\
        \bottomrule
    \end{tabular}%
    \begin{tablenotes}
      \item $^\ddagger$ $\mathit{AVG}(t) = \frac{\sum_{j=1}^t  C(j) -\min\big\{L(j), C(j-1)\big\}}{\sum_{j=1}^t \max \big\{H(j), C(j)\big\} -\min \big\{L(j), C(j)\big\}}$; $^\dagger$ $\mathit{+DM_{\tau}}(t) = \sum \limits_{j=t-\tau+1}^{t} \frac{H(j) -H(j-1)}{\tau}$; $\mathit{-DM_{\tau}}(t) = \sum \limits_{j=t-\tau+1}^{t} \frac{L(j)-L(j-1)}{\tau}$
      \item $^\dagger$ $TRS_{\tau}(t) = \frac{1}{\tau} \sum \limits_{j=t-\tau+1}^{t} TR(j), \quad \textit{with,} \quad TR(j) = \max\Big\{ H(j)-L(j), \ H(j)-C(j-1), \ L(j)-C(j-1) \Big\}$
      \item $^{\dagger\dagger}$ $C(j)$, $H(j)$ and $L(j)$ respectively give the closing,  the highest and the lowest price of day$-j$; $\mathit{HH}_\tau(t)\leftarrow$ the highest high price in the previous $(t-\tau)$ days; $\mathit{LL}_\tau(t)\leftarrow$ the lowest low price in the previous $(t-\tau)$ days; $\mathit{UD}_\tau(t)\leftarrow$ \textit{upward days} during $(t-\tau)$ days; $\mathit{DD}_\tau(t) \leftarrow$ \textit{downward days} in during $(t-\tau)$; $\mathit{UPC}_\tau(t)\leftarrow$ cumulative closing values on upward days during $(t-\tau)$; $\mathit{DPC}_\tau(t)\leftarrow$ the cumulative closing values on downward days during $(t-\tau)$; $\mathit{TV}_\tau(t)\leftarrow$ the volume summation over $(t-\tau)$; $UV_\tau(t)\leftarrow$ cumulative volume restricted to upward days; $\mathit{DV}_\tau(t)\leftarrow$ cumulative volume on downward days
    \end{tablenotes}
  \end{threeparttable}
 \end{adjustbox}
 \end{table*}
 
For the sake of simplicity, let $x_i$ denote the $i^{th}-$feature, which is determined from a particular technical indicator in Table~\ref{table:technical_indicators}; then the labeled learning data $\mathcal{D}$ can be represented as: 
\begin{equation}
    \label{eq:pattern}
    \mathcal{D} = \begin{Bmatrix} \left(x_1, x_2, \dots, x_{n_f}, y\right) ^{(k)} \ \mid \ y \in \{0,1\} \end{Bmatrix}, \qquad k = 1,2,\dots \mathcal{N}
\end{equation}
where, $n_f$ and $\mathcal{N}$ respectively denote the total number of features and patterns. A total of 68 features are obtained by evaluating the technical indicator expressions in Table~\ref{table:technical_indicators}, of which some are parameterized over different time periods $\tau$, \textit{i.e.}, $n_f = 68$. A sliding window of size $w=(\tau+1)$ is used on the daily time series of fundamental trading quantities to extract each pattern in $\mathcal{D}$. For any given day$-t$, the features ($x_1,\dots, x_{n_f}$) are evaluated using the trading information over the period of $[t-\tau, t]$ days, and the corresponding day-ahead prediction label is generated using the closing value of the next day $(t+1)$, see~(\ref{eq:Ypred}). Accordingly, each pattern corresponds to a particular frame of sliding window $(t-\tau+1)$, where $t=1,2, \dots ,N_{days}$ and $N_{days}$ denotes a \textit{number of trading days}. It is emphasized that the prediction is truly \emph{ex-ante} as any information from the prediction day, say $(t+1)$, is not used to extract input features. 

\subsection{Dataset: Pre- and Peri-COVID Stock Index Movements}
\label{subsec:Nasdaq_COVID}

This study focuses on the historical stock index time-series data over four years starting from January, 2017 to May, 2021. The motivation behind the selection of this timeline lies in the fact that it includes stock market behavior approximately two years prior to and after the breakout of the COVID-19 pandemic. The profound impact of the COVID-19 epidemic on the world economy has exacted measures and efforts by policymakers, managers, and academics, and has suffused the stock markets with high volatility trends, as analyzed in~\cite{Gao_EtAl_Covid19_volatility_US_China_2022,Rahman_EtAL_COVID19_volatility_2022,Li_EtAl_Covid19_fear_volatility_2022,BuszkoEtAl2021}. The risk-laden repercussion for financial institutions and investors motivate the development of forecasting models under the changed market behavior. Earlier investigation~\cite{ChandraEtAl2021,Hafiz:Broekaert:2021} on this time period indicated that training data reflecting distinct market behavior prior to the COVID-19 pandemic could have a detrimental effect on the forecasting performance in within-COVID 19 time period. This period, hence, can be thought of as a real-life benchmark to design forecasting models in an environment following a market disruption, with possible inconsistencies in market behaviors prior to and after the disruption. In particular, the historical data in this study is segmented into \textit{pre-} and \textit{within-COVID} time periods as follows: Pre-COVID data ($\mathcal{D}_{pr}$): from January, 2017 to December, 2018; Within-COVID \textit{training} data ($\mathcal{D}_{train}$): from January, 2019 to August, 2020; Within-COVID \textit{testing} data ($\mathcal{D}_{test}$): from August, 2020 to January, 2021; Within-COVID \textit{hold-out} data ($\mathcal{D}_{hold}$): from January, 2021 to May, 2021. 

Further, one of the major concerns in financial forecasting is the issue of inadvertent \emph{data snooping}~\cite{Sullivan_EtAl_data_snooping_technical_trading_1999,Henrique_EtAl_Review_Market_prediction_snooping_2019}, which can lead to irregularly inflated estimation of generalization capabilities. The issue of data snooping is addressed by sequestering data over four time windows, as discussed earlier. Each dataset is kept strictly separate, and their roles are defined as follows: (1) $\mathcal{D}_{train}$: serves as the \emph{training} dataset and is used for estimation of neural network weights (2) $\mathcal{D}_{pr}$ and $\mathcal{D}_{test}$: serve as \emph{test} datasets to evaluate performance over \emph{pre-} and \emph{within-}COVID periods during the optimization of neural architecture. 
(3) $\mathcal{D}_{hold}$: serves as the \emph{hold-out} dataset. This is a truly \emph{out-of-sample} dataset as it is not used in any step of model development, including neural architecture optimization and weight estimation. Accordingly, this dataset is used to assess the generalization capabilities of the identified models.

\subsection{Performance Metrics}
\label{subsec:metrics}

Given that the movement forecasting model is essentially a binary classifier, the \textit{overall classification accuracy} is equivalent to the well-known financial metric `Hit-Rate', which also evaluates the ratio of correct predictions over total predictions~\cite{Shynkevich:McGinnity:2017}. Note that performance assessment using only overall \emph{accuracy} may be misleading as a long-term drift in the historical trading data may lead to a \emph{brute} classifier (\textit{predicting only majority movement in a dataset})~\cite{UlHaq2021,Hafiz:Broekaert:2021}. This study, therefore, considers additional metrics for a balanced assessment of classification performance over both \emph{upwards} and \emph{downwards} movements: the Matthews Correlation Coefficient (MCC), the Balanced Accuracy (BA), and Balanced Error = (1 - BA), see~\cite{Grandini:Bagli:2020} for details.

\section{Multi-objective Co-evolution of Neural Architecture and Features}
\label{sec:Multi-objective_and_Co-evolution_ANN_search}

This study pursues the neural topology as a set of choices for the design of hidden layers (\textit{i.e.}, \textit{size} and \textit{depth}). In addition, the selection of activation function is also being considered. The motivation for this inclusion lies in the earlier investigations of~\cite{Hagg:Mensing:2017}, which showed that an independent selection of the activation function for each neuron can lead to relatively sparse topologies. Accordingly, a candidate neural topology ($\mathcal{T}$) can be represented by a set of tuples, 
\begin{equation}
	\label{eq:nntuple}
   \mathcal{T} \leftarrow \begin{Bmatrix} (s^1,f^1), (s^2,f^2), \dots, (s^{n_\ell}, f^{n_\ell}) \ \Big |\\ \ s^k \in [0, s^{max}], \quad f^k \in \mathcal{F}, \quad \forall{k} \in [1,n_\ell] \end{Bmatrix}
\end{equation}
where, the $i^{th}$-tuple, $(s^i,f^i)$, represents the design choices for the $i^{th}$ hidden layer; $s$ gives the \textit{size} or the number of hidden neurons; $f$ denotes a particular activation function which is selected from the set of activation functions, $\mathcal{F}$, for each hidden layer; $s^{max}$ and $n_\ell$ respectively give the maximum number of hidden neurons and layers, which are user-defined parameters. Such topologies can be categorized as a \textit{semi-heterogeneous}, see Hagg \textit{et al.}~\cite{Hagg:Mensing:2017}. 

Further, from the topological perspective, the selection of features can be viewed as the selection of input neurons. The removal of \textit{irrelevant} and \textit{redundant} features through feature selection can, therefore, be considered a part of the neural design. Before we discuss the co-evolution of features with neural topology, consider the feature selection process in the context of a given set of full features, $X_{\rm full}$:
\begin{align}
    \label{eq:fs}
     X^{\star} & = \Big\{ X \subset X_{\rm full} \ \big{|} \    J(X) = \min \limits_{\forall{X_i} \subset X_{\rm full}} J(X_i) \Big\}, \qquad \text{where,} \quad X_{\rm full} = \begin{Bmatrix} x_1, & x_2, & \dots, & x_{n_f} \end{Bmatrix}
\end{align}
where, $x_i$ and $n_f$ denote the $i^{th}$ feature and the total number of features, respectively; $X \subset X_{\rm full}$ is a candidate feature subset; and $J(\cdot)$ is a suitable criterion function which measures the utility of feature subsets, \textit{e.g.}, classification error. 
Feature selection is a combinatorial problem due to feature correlations, which becomes NP-Hard even for a moderate number of features~\cite{Guyon:Isabelle:2003,Hafiz:Swain:2018}. Usually, the selection of features and the design of neural topology are carried out independently, as discussed in Section~\ref{sec:literature_review}. This study, in contrast, proposes a co-evolution of feature selection with the neural topology, as follows:
\begin{equation}
    \label{eq:ENS}
    \mathcal{A}^{\star} = \argmin \limits_{\mathcal{A}_i \in \Omega}
    \begin{cases}
    \mathcal{E}(\mathcal{A}_i, \mathcal{D}_{test})\\
    \mathcal{C}(\mathcal{A}_i)
    \end{cases}, \ \text{where,} \quad \mathcal{A}_i = \begin{Bmatrix} \mathcal{T}_i, & {X}_i \end{Bmatrix}
\end{equation}
$\mathcal{A}$ denotes an extended neural architecture which is a combination of feature subset $X$ and hidden layer topology $\mathcal{T}$; $\mathcal{D}_{test}$ denotes a test data; and $\mathcal{E}(\cdot)$ and $\mathcal{C}(\cdot)$ respectively denote the criteria to measure the classification performance and complexity of a candidate extended architectures (discussed later in Section~\ref{subsec:criteria}). $\Omega$ denotes the search space of the co-evolution problem, and it is given by,
\begin{align}
\label{eq:searchspace}
\Omega & = \Omega_\mathcal{T}  \times \Omega_F, \qquad
\text{where,} \quad \Omega_\mathcal{T} = \Big( [0, s^{max}] \times \mathcal{F} \Big)^{n_\ell}, \quad \Omega_F = \Big\{ X \ | \ {X} \subset {X}_{\rm full} \wedge X \neq \emptyset \Big\} 
\end{align}
$\Omega_\mathcal{T}$ and $\Omega_F$ denote the search space of neural architectures and features, respectively. It is worth noting that the co-evolution is formulated as a bi-objective problem to balance the complexity of neural architectures with their efficacy, as seen in~(\ref{eq:ENS}).

Further, this study, in particular, focuses on the neural design for forecasting stock index movement over the past four years, which includes two distinct market behaviors stemming from the ongoing COVID-19 pandemic, as discussed in Section~\ref{subsec:Nasdaq_COVID}. The previous investigations~\cite{ChandraEtAl2021,Hafiz:Broekaert:2021} showed that these behavioral changes may be contradictory. The co-evolution problem is, therefore, re-formulated as a multi-objective problem to accommodate distinct market behaviors prior to and during the COVID-19 pandemic, as follows:
\begin{equation}
    \label{eq:nasstock}
    \mathcal{A}^{\star} = \argmin \limits_{\mathcal{A}_i \in \Omega}
    \begin{cases}
    \mathcal{E}_{cv}(\mathcal{A}_i)\\
    \mathcal{C}(\mathcal{A}_i)\\
    \mathcal{E}_{pr}(\mathcal{A}_i)
    \end{cases}
\end{equation}
where, $\mathcal{E}_{pr}(\mathcal{A}_i)$ and $\mathcal{E}_{cv}(\mathcal{A}_i)$ respectively denote the values of balanced error obtained with the architecture $\mathcal{A}_i$ over \emph{pre-COVID} ($\mathcal{D}_{pr}$) and \emph{within-COVID} ($\mathcal{D}_{test}$) datasets, see Section~\ref{subsec:Nasdaq_COVID}.

\section{Search Framework for Co-evolution Approach}
\label{sec:Multi-objective_search}

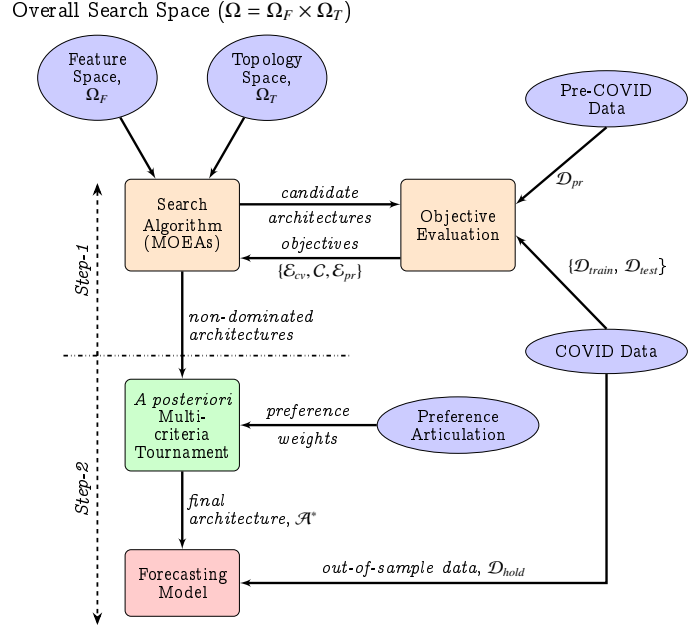
\begin{figure}[!t]
\centering
\begin{adjustbox}{max width=0.55\textwidth}
\tikzstyle{block} = [rectangle, draw, fill=white, 
    text width=7em, text centered, rounded corners, minimum height=4.5em,line width=0.02cm]
\tikzstyle{block2} = [rectangle, draw, fill=white, 
    text width=7em, text centered, rounded corners, minimum height=6.2em,line width=0.02cm]
\tikzstyle{line} = [draw, -latex']
\tikzstyle{cloud} = [draw, ellipse,fill=white, node distance=4.8cm, minimum height=3em, line width=0.02cm,text width=7em,text centered]
\tikzstyle{cloud2} = [draw, ellipse,fill=white, node distance=4.8cm, minimum height=3em, line width=0.02cm,text width=5em,text centered]
\begin{tikzpicture}[auto]
    \centering
   \node [cloud2, fill=blue!20] at (-2,6.5) (FS) {\large Feature Space, $\Omega_F$};
   \node [cloud2, fill=blue!20] at (2,6.5) (TS) {\large Topology Space, $\Omega_T$};
   \node[fit=(FS)(TS),label={[node font=\Large,black]above:Overall Search Space ($\Omega= \Omega_F \times \Omega_T$)}](F){};
   \node [block2, fill=orange!20] at (0,3) (search) {\large Search\\Algorithm\\ (MOEAs)}; 
   \node [block2, fill=orange!20] at (6.5,3) (OE) {\large Objective Evaluation};
   \node[cloud,fill=blue!20] at (10,0) (data) {\large COVID Data}; 
   \node[cloud,fill=blue!20] at (10,6) (predata) {\large Pre-COVID Data}; 
   \draw[dash dot dot, very thick] (-2.8,-0.1) -- (4,-0.1);
   \draw [-{Stealth[length=3mm]},dashed,very thick] (-2,-0.1) --node[rotate=90,above]{\large \emph{Step-1}} (-2,4);
   \draw [dashed,very thick,-{Stealth[length=3mm]}] (-2,0) --node[rotate=90,above]{\large \emph{Step-2}} (-2,-6.5);
   \node[block2,fill=green!20] at (0,-1.75) (MCDM){\large \textit{A posteriori} Multi-criteria Tournament};
   \node[cloud, fill=blue!20] at (6.5,-1.75) (prefArt){\large Preference Articulation}; 
   \node[block,fill=red!20] at (0,-5.5) (final){\large Forecasting Model};
    \path [line,ultra thick] (FS) -- (search);
    \path [line,ultra thick] (TS) -- (search);
    \path [line,ultra thick] (search) --node [right,text width=3.5cm]{\large \emph{non-dominated architectures}} (MCDM);
    \path [line,ultra thick] (MCDM) --node[text width=4cm]{\large \emph{final}\\ \large \emph{architecture}, $\mathcal{A^\ast}$} (final);
    \path [line,ultra thick] (data.90) --node [above]{\large $ \ \qquad \qquad \qquad \{\mathcal{D}_{train}$, $\mathcal{D}_{test}$\}}  (OE.350);
    \path [line,ultra thick] (predata.270) --node [below]{\large $ \quad \mathcal{D}_{pr}$} (OE.20);
    \path [line, ultra thick] (prefArt) --node [above]{\large \textit{preference}} node [below]{\large \textit{weights}} (MCDM); 
    \path [line, ultra thick] (search.20) --node [above]{\large\textit{candidate}} node [below]{\large \textit{architectures}} (OE.160); 
    \path [line, ultra thick] (OE.210) --node [above]{\large\textit{objectives}} node[below] {\large $\{\mathcal{E}_{cv}, \mathcal{C}, \mathcal{E}_{pr} \}$}(search.330); 
    \path [line,ultra thick] (data.270) |-node [above, near end]{\large \emph{out-of-sample data}, $\mathcal{D}_{hold}$} node[below, near end] {} (final.0);
\end{tikzpicture}  
\end{adjustbox}
\caption{Proposed multi-objective co-evolution framework for simultaneous optimization of features and neural topology. APS denotes the set of non-dominated neural architectures identified by the search algorithm.}
\label{f:framework}
\end{figure}

Fig.~\ref{f:framework} shows the overall framework of the proposed co-evolution approach to identify neural architectures under the multi-dataset learning scenario involving \emph{pre-} and \emph{within-}COVID stock market behaviors. This framework can broadly be categorized into two steps: (1) the search for \emph{non-dominated} neural architectures and (2) the selection of final architecture as per the preferences of a Decision Maker (DM). These steps are discussed briefly in the following:

An effective search strategy is crucial to identifying promising combinations of feature subset and neural topology in the extended search space ($\Omega$). To this end, this study considers the two well-known Multi-Objective Evolutionary Algorithms (MOEAs). In essence, MOEAs sample the search space ($\Omega$) to generate and evaluate candidate neural architectures ($\mathcal{A}$) throughout the search process, as seen in Fig.~\ref{f:framework}. A set of \emph{non-dominated} neural architectures is identified at the end of this search process. The details associated with the search process will be discussed in Section~\ref{subsec:MOEAs}.

It is worth noting that non-dominated architectures identified by MOEAs are directly incomparable as they represent a varying degree of trade-offs across the search objectives (\textit{i.e.}, architecture complexity, \emph{pre-} and \emph{within-}COVID classification performance). From the perspective of the DM, such non-dominated architectures represent a set of forecasting models with distinct capabilities. Accordingly, preferences of DM about search objectives can be used to compare and, thus, select from the identified architectures. The second step (see Fig.~\ref{f:framework}) of the proposed framework is designed following these notions, see~Section~\ref{subsec:preferences}.

\subsection{Multi-objective Evolutionary Search Algorithms}
\label{subsec:MOEAs}

Multi-Objective Evolutionary Algorithms (MOEAs) aim to approximate the true Pareto front (PF) of the given problem as accurately as possible. This requires that the Approximated Pareto Front (APF) includes non-dominated solutions which are not only close to the PF (\textit{i.e.}, \textit{convergence}) but are also well distributed over the entire PF (\textit{i.e., diversity}). The balance of convergence with the diversity of the APF is crucial for well-informed \emph{a posteriori} decision making. The diversity preserving mechanisms in most MOEAs can be categorized into dominance-based or decomposition-based perspectives, \textit{see}~\cite{Cai:Li:2015} \textit{for details}. To accommodate both perspectives, two distinct MOEAs are considered: (1) an augmented version of the classical dominance-based algorithm, NSGA-II~\cite{Deb:Pratap:2002,Ishibuchi:Tsukamoto:2010} (2) External Archive Guided MOEA Based on Decomposition (EAGD)~\cite{Cai:Li:2015}.

This study selects the classical NSGA-II~\cite{Deb:Pratap:2002} from dominance-based MOEAs due to its popularity. It is worth noting that while the NSGA-II includes the \textit{crowding distance operator} to improve the solution diversity, it has been shown that the diversity can further be improved by replacing the conventional recombination operators in NSGA-II (\textit{e.g.}, \textit{single-point} or \textit{uniform crossover}) with a \textit{non-geometric crossover} operator~\cite{Ishibuchi:Tsukamoto:2010,Hafiz:Swain:MOEA:2020}, which is adopted here. The design of the non-geometric operator, its parameter settings, and their impacts have been investigated in detail by Ishibuchi \textit{et al.} in~\cite{Ishibuchi:Tsukamoto:2010} and, therefore, are not discussed here. The other implementation details, such as \textit{non-dominated sorting} and \textit{crowding tournament selection operator}, are implemented following the original proposal of Deb et al.~\cite{Deb:Pratap:2002}. Further, the decomposition-based MOEAs often under-perform on non-continuous Pareto fronts, which are typically associated with combinatorial problems similar to feature selection~\cite{Cai:Li:2015,Hafiz:Swain:MOEA:2020}. EAGD~\cite{Cai:Li:2015} overcomes this issue by a hybrid \emph{dominance-decomposition} based approach for multi-objective combinatorial problems and, therefore, is considered as the second search algorithm.

As discussed earlier, both NSGA-II and EAGD generate and evaluate candidate neural architectures by sampling the search space $\Omega$. To this end, both algorithms encode candidate neural architectures as an $n-$dimensional binary string, which will be discussed in Section~\ref{subsec:encoding}. The efficacy of each candidate architecture is evaluated across all the search objectives, \textit{i.e.}, architectural complexity, and classification performance over \emph{pre-} and \emph{within-}COVID datasets. This evaluation procedure will be discussed in Section~\ref{subsec:criteria}.

\subsubsection{Binary Encoding}
\label{subsec:encoding}

Each candidate neural architecture, $\mathcal{A} = \{ X, \ \mathcal{T} \}$, is encoded by an $n-$dimensional binary string, where, $n=n_f + \ (n_\ell \cdot n_{bits})$. The first $n_f-$bits of such string encode the feature subset ($X$), whereas the remaining bits of the string encode the neural topology ($\mathcal{T}$). A typical binary encoding (denoted as $\mathcal{B}$) is given by,

\begin{align} 
\label{eq:beta}
    & \mathcal{B} = \left[ \overbrace{\begin{matrix} \beta_{1}, & \dots & \beta_{n_f} \end{matrix}}^{features} \ \ \overbrace{\begin{matrix} \beta_{n_f+1}, & \dots \beta_{n_f+n_{bits}}, & \beta_{n_f+(n_{bits}+1)}, & \dots, \beta_n \end{matrix}}^{topology} \right], \qquad
     \text{where,} \quad n = (n_{bits} \cdot n_\ell) \ + \ n_f
\end{align}

The feature subset $X$ is encoded by a binary substring of length $n_f$. Each bit (denoted by $\beta$) of such a substring encodes whether the corresponding feature is included in $X$, \textit{e.g.}, $\beta_j=1$ indicates that the $j^{th}-$feature is included in $X$, $x_j \in X$. Further, each hidden layer of $\mathcal{T}$ is encoded by an $n_{bits}-$tuple. The first $(n_{bits}-1)$ bits of such tuple encode the \textit{size} (number of hidden neurons, $s$), whereas the last bit encodes the selection of an activation function, $f\in\mathcal{F}$. The set of possible activation functions is given by, $\mathcal{F}=\begin{Bmatrix} sigmoid, \ tanh \end{Bmatrix}$. A total of $(n_\ell \cdot n_{bits})$ bits is required to encode $\mathcal{T}$, as the maximum number of hidden layers is limited to $n_\ell$.  

\begin{algorithm}[!t]
    \small
    \SetKwInOut{Input}{Input}
    \SetKwInOut{Output}{Output}
    \SetKwComment{Comment}{*/ \ }{}
    \Input{Search Agent, $\mathcal{B}_i = \begin{Bmatrix} \beta_{i,1}, & \dots, & \beta_{i,n} \end{Bmatrix}$}
    \Output{Criterion Function, $\vec{J}(\mathcal{A}_i) \leftarrow \Big[ \mathcal{E}_{cv}(\mathcal{A}_i),  \quad \mathcal{C}(\mathcal{A}_i), \quad \mathcal{E}_{pr}(\mathcal{A}_i) \Big]^T$ }
    \algorithmfootnote{ ${\rm L}(\cdot)$ denotes \emph{loss-function} for the weight estimation~\cite{Moller:1993}}
    \BlankLine
    \Comment*[h] {Decode the feature subset, ${\rm X}_i$}\\
    \BlankLine
    ${\rm X}_i \leftarrow \emptyset$
    
    \For{j = 1 to $n_f$} 
         { 
           \uIf{$\beta_{i,j}=1$}
           {${\rm X}_i \leftarrow \{ {\rm X}_i \cup x_j \}$  \Comment*[h] {add the $j^{th}$ feature}
           }
        }
    \BlankLine
    $\mathcal{B}_i \leftarrow \mathcal{B}_i \setminus \begin{Bmatrix} \beta_{i,1}, & \beta_{i,2}, & \dots, & \beta_{i,n_{f}} \end{Bmatrix}$
    \BlankLine
    \Comment*[h] {Decode the neural topology, $\mathcal{T}_i$}\\
    \BlankLine
    $\mathcal{T}_i \leftarrow \emptyset$\nllabel{l:crf1}

    \For{j = 1 to $n_\ell$} 
        { \BlankLine
           \Comment*[h] {Layer Size}
           
           $\displaystyle s_j \leftarrow \sum \limits_{p=1}^{(n_{bits}-1)} \beta_{i,p} \times 2^{(n_{bits} -p -1)}$ 
           \BlankLine
           \Comment*[h] {Activation Function}
           
           \lIf{$\beta_{i,n_{bits}}$ = 1}
           { $f^j \leftarrow \textit{sigmoid} $}
           \lElse{$f^j \leftarrow \tanh $}
           \BlankLine
           $\mathcal{T}_i \leftarrow \left\{\mathcal{T}_i \cup (s^j,f^j) \right\}$\\
           $\mathcal{B}_i \leftarrow \mathcal{B}_i \setminus \begin{Bmatrix} \beta_{i,1}, & \dots, & \beta_{i,n_{bits}} \end{Bmatrix}$
        } 
    \BlankLine \nllabel{l:crf2}
    $\mathcal{A}_i \leftarrow \begin{Bmatrix} \mathcal{T}_i, \rm{X}_i \end{Bmatrix}$ \Comment*[h] {Candidate Architecture} \nllabel{l:crf3}\\
    \BlankLine
    \Comment*[h] {Architecture Complexity}\\
    $\mathcal{C}(\mathcal{A}_i) \leftarrow \frac{1}{3} \left\{ \frac{\big| X_i \big|}{n_f} + \frac{\Big|\begin{Bmatrix} (s^k,f^k) \ | \ s^k \neq 0, \forall k \in [1,n_\ell] \end{Bmatrix}\Big|}{n_\ell} + \sum \limits_{k=1}^{n_\ell} \frac{s^k}{s^{max}} \right\}$\nllabel{l:crf4}\\
    \BlankLine
    \Comment*[h] {Estimation of Network Performance}\\
    \BlankLine
    \For{k = 1 to $cycles$ \nllabel{l:crf41}}
    {
        \BlankLine
        \Comment*[h] {Weight Estimation, see~\citep{Moller:1993}}\\
        \BlankLine
        $\mathcal{W}_{i,k}^{\ast} \leftarrow \argmin \limits_{\mathcal{W}} \ {\rm L}(\mathcal{A}_i,\mathcal{W},\mathcal{D}_{train})$
        \BlankLine
        Determine balanced errors: $\mathcal{E}_k(\mathcal{A}_i,\mathcal{W}_{i,k}^{\ast},\mathcal{D}_{pr})$ and $\mathcal{E}_k(\mathcal{A}_i,\mathcal{W}_{i,k}^{\ast},\mathcal{D}_{test})$\\
    }
    \BlankLine   
    \Comment*[h] {Efficacy over COVID-period}\\
    $\mathcal{E}_{cv}(\mathcal{A}_i) \leftarrow \displaystyle\frac{1}{cycles} \sum \limits_{k=1}^{cycles} \mathcal{E}_k(\mathcal{A}_i,\mathcal{W}_{i,k}^{\ast},\mathcal{D}_{test})$, 
    \BlankLine
    \Comment*[h] {Efficacy over Pre-COVID period}\\
    $\mathcal{E}_{pr}(\mathcal{A}_i) \leftarrow \displaystyle\frac{1}{cycles} \sum \limits_{k=1}^{cycles} \mathcal{E}_k(\mathcal{A}_i,\mathcal{W}_{i,k}^{\ast},\mathcal{D}_{pr})$\nllabel{l:crf5}
\caption{\small Criteria Evaluation, $J(\cdotp)$}
\label{al:crf}
\end{algorithm}
\subsubsection{Criteria Evaluation}
\label{subsec:criteria}

Throughout the search process, each candidate neural architecture is evaluated following the steps outlined in Algorithm~\ref{al:crf}. The process begins by decoding a binary string ${\mathcal{B}}_i$, which represents an $i^{th}-$candidate architecture in both NSGA-II and EAGD, following the steps in Line~\ref{l:crf1}~-~\ref{l:crf3}, Algorithm~\ref{al:crf}. Its performance is determined next in terms of the main three search objectives, as outlined in Line~\ref{l:crf4}~-~\ref{l:crf5}, Algorithm~\ref{al:crf}, and discussed in detail in the following:

In this study, the complexity of neural architectures is defined directly as a function of different topological components as well as features,

\begin{equation}
   \label{eq:complexity2}
   \begin{aligned}
   \mathcal{C}(\mathcal{A}_i) = \frac{1}{3} \Bigg\{ \frac{\big| X_i \big|}{n_f} + \frac{\Big|\begin{Bmatrix} (s^k,f^k) \ | \ s^k \neq 0, & \forall k \in [1,n_\ell] \end{Bmatrix}\Big|}{n_\ell} + \sum \limits_{k=1}^{n_\ell} \frac{s^k}{s^{max}} \Bigg\} 
   \end{aligned}
\end{equation}

where, $\mathcal{A}_i$ denotes a candidate architecture; $|\cdot|$ determines the cardinality of the given set; $X_i$ denotes the feature subset encoded by $\mathcal{A}_i$; $n_f$ gives the total number of features; $n_\ell$ and $s^{max}$ respectively give the maximum number of hidden layers and neurons, respectively. The function, $\mathcal{C}$, is bounded in $[0,1]$ and will attain the maximum value for the \textit{mother architecture}, \textit{i.e.}, the architecture which includes all features, the maximum allowable neurons, and layers. $\mathcal{C}$ thus measures the complexity of any given architecture relative to the \textit{mother architecture}. It is easy to follow that a lower value of this function is desirable.

Next, we focus on the criteria functions that determine the classification performance. This process begins by estimating the network weights ($\mathcal{W}$) for the candidate architecture using the scaled-conjugate gradient descent algorithm~\cite{Moller:1993}. It is worth emphasizing that the weight estimation is often influenced by various factors including but not limited to \textit{local minima} and \textit{weight initialization}, which may translate into an incorrect estimation of the architecture's efficacy~\cite{Yao:1999}. To minimize such effects, for the given architecture, the weight estimation is repeated over multiple learning cycles, as outlined in Line~\ref{l:crf41}~-~\ref{l:crf5}, Algorithm~\ref{al:crf}. The subsequent average values of balanced error over \emph{pre-} and \emph{within-}COVID datasets, $\mathcal{E}_{cv}(\mathcal{A}_i)$ and $\mathcal{E}_{pr}(\mathcal{A}_i)$, serve as two search objectives in addition to the aforementioned architectural complexity, $\mathcal{C}$.
 
 \subsection{Preference Articulation}
\label{subsec:preferences}

The non-dominated architectures identified for the co-evolution problem in~(\ref{eq:nasstock}) by MOEAs are incomparable. Hence, the \textit{a posteriori} selection of the final architecture depends on the stated \textit{preferences} of the decision maker (DM), \textit{i.e.}, the relative importance of the three different objectives from the perspective of the DM. This procedure is outlined in Algorithm~\ref{al:mtd} and is discussed in detail in the following: Let $\Gamma$ denote the Approximate Pareto Set (APS), which represents a set of non-dominated extended architectures identified by a particular MOEA for the co-evolution problem in~(\ref{eq:nasstock}), \textit{i.e.},
\begin{equation}
    \Gamma =\begin{Bmatrix} \mathcal{A}_{1}, & \mathcal{A}_{2}, & \dots \end{Bmatrix}, \qquad \Lambda =\begin{Bmatrix} \vec{J}(\mathcal{A}_{1}), & \vec{J}(\mathcal{A}_{2}), & \dots \end{Bmatrix}
\end{equation}
where, \begin{footnotesize}$\vec{J}(\mathcal{A}_i) = \Big[ \mathcal{E}_{cv}(\mathcal{A}_i),  \quad \mathcal{C}(\mathcal{A}_i), \quad \mathcal{E}_{pr}(\mathcal{A}_i) \Big]^T$\end{footnotesize}, with $i = 1, 2, \dots, |\Gamma|$; $\Lambda$ denotes the Approximate Pareto Front (APF) corresponding to $\Gamma$. Each identified non-dominated architecture $\mathcal{A}_i \in \Gamma$ represents a distinct degree of trade-off across \textit{complexity} and \textit{forecasting performance} in \textit{pre-} and \textit{within-COVID} periods.

The crucial first step is to quantify often \textit{abstract} and \textit{partial} preferences of the DM. This study relies on the multiplicative preference relations~\cite{Zhang:Chen:2004} for this purpose, which translates preferences into quantifiable weights (\textit{denoted by} $\theta$) for each objective. To this end, the quantification process starts by obtaining the preferences from the DM in terms of an ordered ranking ($O$) for each objective, in the decreasing order of their importance. In the next step, the \textit{intensity} of the objective ranking (denoted by $\mathcal{I}$) is selected on a scale from `$1$' (\textit{indifference}) to `$9$' (\textit{extreme prejudice}). To understand this further, let the objective rankings and the preference intensity specified by the DM be given by:

\begin{equation}
    \label{eq:prefExample1}
    O = \begin{bmatrix} O_{cv} & O_{\mathcal{C}} & O_{pr}\end{bmatrix} = \begin{bmatrix} 1 & 2 & 3 \end{bmatrix}, \text{and \ } \mathcal{I}=9  
\end{equation}

where, $O_{cv}$, $O_{\mathcal{C}}$, and $O_{pr}$ respectively denote ranking for \textit{within-COVID} performance, \textit{complexity} and \textit{pre-COVID} performance. This objective ranking indicates that the highest preference is given to \textit{within-COVID} performance, followed by \textit{complexity} and \textit{pre-COVID} performance. For these specifications, the preference relations ($\pi$) between the objectives and, ultimately, the preference weights ($\theta$) are determined by following the steps outlined in Line~\ref{line:prefw1}~-~\ref{line:prefw2}, Algorithm~\ref{al:mtd}, as follows:

\begin{equation}
\begin{aligned}
    \label{eq:prefExample2}
    & \begin{bmatrix} 
        \pi_{1,1} & \pi_{1,2} & \pi_{1,3}\\ \pi_{2,1} & \pi_{2,2} & \pi_{2,3}\\
        \pi_{3,1} & \pi_{3,2} & \pi_{ 3,3} \end{bmatrix} 
        = \begin{bmatrix} 1 & \sqrt{9} & 9\\ \frac{1}{\sqrt{9}}  & 1 & \sqrt{9}\\ \frac{1}{9} & \frac{1}{\sqrt{9}} & 1 \end{bmatrix}, \quad \text{which yields,}\\
        & \vec{\theta} = \frac{\begin{bmatrix} \theta_1 & \theta_2 & \theta_3\end{bmatrix}}{\sum \theta} = \frac{\begin{bmatrix} 3 & 1 & 0.33 \end{bmatrix}}{4.33} = \begin{bmatrix} 0.69 & 0.23 & 0.07\end{bmatrix}
\end{aligned}
\end{equation}

In the last step of the \textit{a posteriori} selection, the preference weights are used to select the final architecture using the Multi-Criteria Tournament Decision (MTD)~\cite{Parreiras:Vasconcelos:2009}. The rationale of this approach is to rank each architecture, $\mathcal{A}_i \in \Gamma$, over a particular objective$-p$ through a tournament comparison with the set of remaining non-dominated architectures, \textit{i.e.}, $\{\Gamma\setminus\mathcal{A}_i$\}. This process is shown in Line~\ref{line:mtd2}~-~\ref{line:mtd3}, Algorithm~\ref{al:mtd}. In particular, the tournament wins ($\tau$) and tournament function ($\Phi$) over all search objectives are determined first. This is followed by the evaluation of the \textit{global rank} ($\mathcal{R}$) which is a function of the \textit{tournament wins} ($\Phi$) and the \textit{preference weight} ($\theta$), see~Line~\ref{line:mtd4}, Algorithm~\ref{al:mtd}. Finally, the architecture with the maximum global rank is selected as the final architecture, see~Line~\ref{line:mtd6}, Algorithm~\ref{al:mtd}. 

\begin{algorithm}[!t]
    \small
    \SetKwInOut{Input}{Input}
    \SetKwInOut{Output}{Output}
    \SetKwComment{Comment}{*/ }{}
    \Input{Pareto set, $\Gamma^*=\{ \mathcal{A}_1, \mathcal{A}_2, \dots \}$; Pareto front, $\Lambda^*=\{ \vec{J}(\mathcal{A}_1), \vec{J}(\mathcal{A}_2), \dots \}$}
    \Output{Selected Structure, $\mathcal{A}^{*}$}
    \algorithmfootnote{$n_{obj}=3$ denotes total number of objectives; Criteria Function: $\vec{J}(\mathcal{A}_i) = \Big[ J_1(\mathcal{A}_i), \ J_2(\mathcal{A}_i), \ J_3(\mathcal{A}_i) \Big]^T$, where, $J_1(\mathcal{A}_i)\leftarrow \mathcal{E}(\mathcal{A}_i,\mathcal{D}_{test})$, $J_2(\mathcal{A}_i) \leftarrow \mathcal{C}(\mathcal{A}_i)$ and $J_3(\mathcal{A}_i) \leftarrow \mathcal{E}(\mathcal{A}_i,\mathcal{D}_{pr})$.}
    \BlankLine
    \Comment*[h] {Preference formulation}\\
     Specify the preference intensity, $\mathcal{I} \in [1,9]$; and the objective rankings, $O = \begin{bmatrix} O_{cv} & O_{\mathcal{C}} & O_{pr} \end{bmatrix}$  \nllabel{line:prefw1}\\
    \BlankLine
     \For{i = 1 to $n_{obj}$} 
     {  
         \BlankLine
         \For{j = 1 to $n_{obj}$}
         {
          $\pi_{i,j} = \mathcal{I}^{\left( \frac{O_j - O_i} {n_{obj}-1}\right)}$ \Comment*[h] {pref. relations}
         } 
         \BlankLine
         $\theta_i = \left( \prod \limits_{j=1}^{n_{obj}} \pi_{i,j} \right)^{1/n_{obj}}$ 
     }
    \BlankLine
    $\vec{\theta} = {\begin{bmatrix} \theta_1 & \theta_2 & \dots & \theta_{n_{obj}} \end{bmatrix}}\bigg{/}{\sum_{p=1}^{n_{obj}} \theta_p}$ \Comment*[h] {weights} \nllabel{line:prefw2}
    \BlankLine
    \BlankLine
    \Comment*[h] {Tournament function}\\
    \BlankLine
    \For{i = 1 to $|\Lambda^*|$\nllabel{line:mtd1}}
    {
        \BlankLine
        \For{p = 1 to $n_{obj}$ \nllabel{line:mtd2}} 
        {
          \BlankLine
          $\tau_{i,p} \leftarrow 0$\\
          \BlankLine
          \For{j = 1 to $|\Lambda^*|$}
          {   \BlankLine
              \lIf{$J_{p}(\mathcal{A}_j)>J_{p}(\mathcal{A}_i)$}
                {$\tau_{i,p} \leftarrow \tau_{i,p} + 1$ \Comment*[h] {tournament wins}}  
          }
          $\Phi_p(\mathcal{A}_i) \leftarrow \frac{\tau_{i,p}}{|\Lambda^*|-1}$ \Comment*[h] {tournament function} \nllabel{line:mtd5}
        }\nllabel{line:mtd3} 
        $\mathcal{R}(\mathcal{A}_i) \leftarrow \left( \prod \limits_{p=1}^{n_{obj}} \Phi_p(\mathcal{A}_i)^{\theta_p} \right)^{\frac{1}{n_{obj}}}$ \Comment*[h] {global rank}\nllabel{line:mtd4}
    }
     \BlankLine
    Select the architecture with the maximum $\mathcal{R}(\cdot)$, \textit{i.e.}, 
     $\mathcal{A}^{*} \leftarrow \left\{ \mathcal{A}_i | \mathcal{R}(\mathcal{A}_i) = \arg \max \mathcal{R}(\mathcal{A}_k), \forall{\mathcal{A}_k} \in \Gamma^* \right\}$\nllabel{line:mtd6}
\caption{\small \textit{A posteriori} selection}
\label{al:mtd}
\end{algorithm}

To understand this procedure, consider an Approximate Pareto front (APF, $\Lambda$) containing four non-dominated architectures ($\mathcal{A}_1, \mathcal{A}_2, \dots, \mathcal{A}_4$) with the objective values given by (\ref{eq:exmObj}). For each architecture in this APF, \emph{tournament wins} ($\tau$) and \emph{tournament function} ($\Phi$) are determined as follows (see~Lines~\ref{line:mtd2}~-~\ref{line:mtd3}, Algorithm~\ref{al:mtd}):
\begin{equation}
\label{eq:exmObj}
\begin{aligned}
\Lambda & = \begin{bNiceMatrix}[first-row,last-row,first-col,last-col]
    & \mathcal{E}_{cv} & \mathcal{C} & \mathcal{E}_{pr}\\
    & 0.43  & 0.27  & 0.46 &  \vec{J}(\mathcal{A}_1)\\
    & 0.42  & 0.30  & 0.48 &  \vec{J}(\mathcal{A}_2)\\
    & 0.41  & 0.36  & 0.47 &  \vec{J}(\mathcal{A}_3) \\
    & 0.45  & 0.65  & 0.45 &  \vec{J}(\mathcal{A}_4)\\
    &     &     &     &   \\
    \end{bNiceMatrix},\quad \text{ gives,} \quad \tau = \begin{bNiceMatrix}[first-row,last-row,first-col,last-col]
    & \mathcal{E}_{cv} & \mathcal{C} & \mathcal{E}_{pr}\\
 & 1     & 3     & 2   &  \mathcal{A}_1\\
 & 2     & 2     & 0   &  \mathcal{A}_2\\
 & 3     & 1     & 1   &  \mathcal{A}_3\\
 & 0     & 0     & 3   &  \mathcal{A}_4\\
 &       &       &     &  \\
\end{bNiceMatrix},\quad 
\text{and,} \quad \Phi = \frac{\tau}{|\Lambda^\ast|-1} = \begin{bNiceMatrix}[first-row,last-row,first-col,last-col]
    & \mathcal{E}_{cv} & \mathcal{C} & \mathcal{E}_{pr}       \\
 & 0.33   & 1     & 0.67  &  \mathcal{A}_1\\
 & 0.67  & 0.67  & 0   &  \mathcal{A}_2\\
 & 1  & 0.33  & 0.33   &  \mathcal{A}_3\\
 & 0     & 0   & 1   &  \mathcal{A}_4\\
 &       &       &     &   \\
\end{bNiceMatrix}
\end{aligned}
\end{equation}
Next, the \emph{global rank} for each architecture in $\Lambda$ is determined by considering the objective rankings, $O$ in~(\ref{eq:prefExample1}) and the consequent preference weights, $\vec{\theta}$ in~(\ref{eq:prefExample2}) as follows:
\begin{equation}
\label{eq:exmObj2}
\mathcal{R} = \begin{bNiceMatrix}[first-row,last-row,first-col,last-col]
&  &        \\
& 0.77  &  \mathcal{A}_1\\
& 0  &  \mathcal{A}_2\\
& 0.89  &  \mathcal{A}_3\\
& 0  &  \mathcal{A}_4\\
&    & \\
\end{bNiceMatrix}
\end{equation}
Here, $\mathcal{A}_3$ is selected as the \emph{preferred} architecture since it has the maximum value of $\mathcal{R}$.

\section{Comparative Evaluation: Baseline Neural Design Approaches}
\label{sec:compeval}

This study considers a total of $21$ different neural architecture design approaches, which are broadly categorized into three \emph{baseline} approaches for comparative evaluation purposes. These baselines represent prevailing neural design approaches in most of the existing stock index movement investigations (see Section~\ref{sec:literature_review}), as will be discussed in the following subsections.

\subsection{Baseline-1: \textit{a priori} Feature Selection and Rule of Thumbs}
\label{sec:baseline1}

The first \emph{baseline} is designed to reflect most neural forecasting models for stock movement, which are designed in two sequential steps: the input dimensionality of the neural network is reduced first through either pre-processing step such as PCA or feature selection \emph{filters}, see Section~\ref{subsec:featsel}. Next, the neural topology is selected either following a \emph{rule-of-thumb} or via trial-and-error (see Section~\ref{subsec:nas}). In particular, the following three input dimensionality reduction methods are being considered for this baseline: PCA, \emph{minimum redundancy-maximum relevancy} (mRmR)~\cite{Peng:Long:2005} and correlation-based feature selection (CFS)~\cite{Hall:1999}. Following the investigations in~\cite{Zhong:Enke:2017,Zhong:Enke:2017b}, a classical PCA variant is applied to the full feature set containing $68$ features. The resultant top 44 principal components explain $>99.99\%$ of data variance. Hence, the subsequently transformed datasets with 44 components are used for neural network training and testing. Further, both feature selection filters (mRmR and CFS) essentially focus on identifying a feature subset that reduces feature-to-feature interaction (\emph{redundancy}) while increasing feature-to-class interaction (\emph{relevance}). We refer to~\cite{Peng:Long:2005,Hall:1999} for implementation details of mRmR and CFS. Both filters use a greedy search approach (best-first search) and identified subsets with $17$ (mRmR) and $29$ (CFS) features.

In the next step of this baseline, the reduced feature subsets are combined with distinct neural topologies, which are designed with six rules-of-thumb given in~\ref{sec:app}. We refer to these \emph{rules-of-thumb} for a systematic design of neural topology to replace empirical trial-and-error topological designs, which is often encountered in the existing studies, \textit{e.g.}, see~\cite{Peng:Albuquerque:2021,UlHaq2021,Liu:Wang:2019,Zhong:Enke:2017b,Zhong:Enke:2017,Inthachot:Boonjing:2016,Asadi2012,Tsai:Hsiao:2010,Lam:2004,Enke:2004,Kim2003,Cao:Tay:2003}.
\subsection{Baseline-2: \textit{a priori} Feature Selection and Topology Optimization}
\label{sec:baseline2}

The empirical design of neural topology via trial-and-error or rules-of-thumb (similar to baseline-1) may not lead to an optimal topology. Therefore, for sake of fair comparison, the second baseline optimizes neural topology after the dimensionality reduction step. The key distinction here, with respect to the proposed co-evolution problem in~(\ref{eq:nasstock}), is that the reduced feature subset (\textit{and therefore the subset of input neurons}) remains fixed during the subsequent neural topology search. The optimization of only topology can be formulated as follows:

\begin{equation}
    \label{eq:baseline}
    \mathcal{T}^{\ast} = \argmin \limits_{\mathcal{T}_i \in \Omega_\mathcal{T}} 
    \begin{cases}
    \mathcal{E}(\mathcal{T}_i,\rm{X}_{d}^\ast, \mathcal{D}_{test})\\
    \mathcal{C}(\mathcal{T}_i,\rm{X}_{d}^\ast)\\
     \mathcal{E}(\mathcal{T}_i,\rm{X}_{d}^\ast, \mathcal{D}_{pr})
    \end{cases}
\end{equation}
where, $\rm{X}_{d}^\ast$ denotes the subset of $d-$features which is identified through \textit{a priori} dimensionality reduction step. The procedure for the baseline topology optimization is similar to the overall process outlined in Section~\ref{sec:Multi-objective_search}, except for one key difference: Given that the features have been selected \textit{a priori}, the candidate solutions in the baseline neural architecture search encode only the design of hidden layers (see Section~\ref{subsec:encoding}), as follows: $\mathcal{B} = \overbrace{\left[ \begin{matrix} \beta_{1}, & \beta_{2}, & \dots, & \beta_{(n_{bits} \cdot n_\ell)} \end{matrix} \right]}^{topology}$

Note that the search space for the neural topology selection reduces to $\Omega_\mathcal{T}$. The caveat, however, is the loss of the degree of freedom to adjust the design of downstream hidden layers according to the feature subset under consideration. A possible trade-off in the performance of the evolved neural topologies thus may be expected.

\subsection{Baseline-3: Scalarized Co-evolution Approach}
\label{subsec:baselineScalerized}

This study considers the \textit{scalarized} multi-criteria approach to co-evolve the feature subset and the neural topology proposed in~\cite{Hafiz:Broekaert:2021} as the third \emph{baseline} search approach, as follows:
\begin{equation}
\begin{aligned}
    \mathcal{A}^{\star} & = \argmin \limits_{        \mathcal{A}_i \in \Omega} \ \mathcal{J}(\mathcal{A}_i), \qquad \textit{where,} \qquad
    \mathcal{J}(\mathcal{A}_i) =  \theta_{\rm{E}} \times {\rm{E}}(\mathcal{A}_i,\mathcal{D}_{test}) \ + \ \theta_\mathcal{C} \times \mathcal{C}(\mathcal{A}_i) \ + \ \mathcal{P}(\mathcal{A}_i) 
\end{aligned}
\end{equation}
where, $\rm{E}(\cdot)$ and $\mathcal{C}(\cdot)$ respectively give the overall \emph{classification error} and the \emph{complexity} of the architecture under consideration; $\theta_{\rm{E}}$ and $\theta_\mathcal{C}$ denote the preferences of the DM; and $\mathcal{P}$ denotes a penalty function which is defined as an $\epsilon-$constraint over both COVID ($\mathcal{D}_{test}$) and pre-COVID data ($\mathcal{D}_{pr}$), as follows: 
\begin{equation}
\label{eq:penalty_epsilon}
\begin{aligned}
    \mathcal{P}(\mathcal{A}_i) =  5 \times \Bigg[ & \max \Big\{ 0, \ \epsilon_1 - \Phi(\mathcal{A}_i,\mathcal{D}_{test}) \Big\} + \max \Big\{ 0, \ \epsilon_2 - \Phi(\mathcal{A}_i,\mathcal{D}_{pr}) \Big\} + \max \Big\{ 0, \ {\rm{E}}(\mathcal{A}_i,\mathcal{D}_{pr}) - \epsilon_3 \Big\} \Bigg] 
\end{aligned}
\end{equation}
where, $\Phi(\cdot)$ denotes Matthews correlation coefficient; $\epsilon_1$, $\epsilon_2$ and $\epsilon_3$ represent the pre-specified thresholds. A detailed discussion on the rationale behind the penalty function as well as the selection of the thresholds can be found in~\citep{Hafiz:Broekaert:2021}. It is worth emphasizing that the goal of both the \textit{proposed} and \textit{scalarized} approach is to co-evolve the feature subset and the neural topology, \textit{i.e.}, they operate in the common problem search space, $\Omega$. However, the search objectives of these approaches differ, especially in terms of how the pre-COVID data, $\mathcal{D}_{pr}$, is being used. In the \textit{proposed} approach, the classification performance over $\mathcal{D}_{pr}$ is one of the main search objectives; see~(\ref{eq:nasstock}) and Section~\ref{sec:Multi-objective_and_Co-evolution_ANN_search}. In contrast, in the \textit{scalarized} approach, the classification performance over $\mathcal{D}_{pr}$ serves as the secondary search objective. This is apparent from the penalty function in~(\ref{eq:penalty_epsilon}), wherein the search is \textit{`guided'} to keep the error over $\mathcal{D}_{pr}$ below a pre-fixed threshold ($\epsilon_3$), and there is no incentive for further reduction, \textit{i.e.}, all neural architectures with $ \rm{E}(\mathcal{X}_i,\mathcal{D}_{pr}) \leq \epsilon_3$ are considered to be equally acceptable when only the performance over $\mathcal{D}_{pr}$ is considered. 

\section{Results}
\label{sec:results}

\subsection{Experimental Setup}
\label{subsec:experimental_search_setup}

The neural architectures are identified for the NASDAQ data over the four years (see Section~\ref{subsec:Nasdaq_COVID}) using both the proposed approach (Section~\ref{sec:Multi-objective_and_Co-evolution_ANN_search} and~\ref{sec:Multi-objective_search}) and the baseline search approaches (Section~\ref{sec:compeval}). The search framework outlined in Section~\ref{sec:Multi-objective_search} is used for the proposed and baseline-2 search environments. A total of 40 independent runs of MOEAs (NSGA-II and EAGD) are carried out, where each run is set to terminate after 15,000 Function Evaluations (FEs). At the end of each run, the identified non-dominated neural architectures and the corresponding approximate Pareto front are recorded. The dominance of such recorded neural architectures is again determined after 40 independent runs, and only non-dominated architectures are considered for further analysis. The search space of neural architectures is determined as follows: A total of $68$ features are considered, which are derived from the technical indicators given in Table~\ref{table:technical_indicators}, \textit{i.e.}, $n_f = 68$. The maximum number of hidden layers is set to 2, \textit{i.e.}, $n_\ell = 2$. Each layer can have the maximum of $128$ neurons, \textit{i.e.}, $s^{max} = 128$. The \emph{mother} or the most complex neural architecture ($\mathcal{C}=1$) for this search space contains all $68$-features as inputs and two hidden layers with $128$ neurons. 

Further, the search parameters of MOEAs are selected \textit{empirically} and based on earlier investigations in~\cite{Cai:Li:2015,Ishibuchi:Tsukamoto:2010,Hafiz:Swain:MOEA:2020}. In particular, the following settings are used for both NSGA-II and EAGD: \textit{population size}$\rightarrow50$; \textit{mutation rate}$\rightarrow1/n$, where $n$ is the total number of search variables (see Section~\ref{subsec:encoding}). The other algorithm-specific parameters are selected as follows: (1) NSGA-II: \textit{crossover rate}$\rightarrow0.9$; \textit{probability of non-geometric crossover}$\rightarrow0.8$; \textit{probability of bit-flip}$\rightarrow1/n$. (2) EAGD: \textit{crossover rate}$\rightarrow1.0$; \textit{learning generations:}$\rightarrow8$; \textit{neighbors:}$\rightarrow10\%$ of population size.

\subsection{Co-evolution: Effects of search algorithm}
\label{subsec:searcheffect}

\begin{SCfigure}
\includegraphics[width=0.32\textwidth]{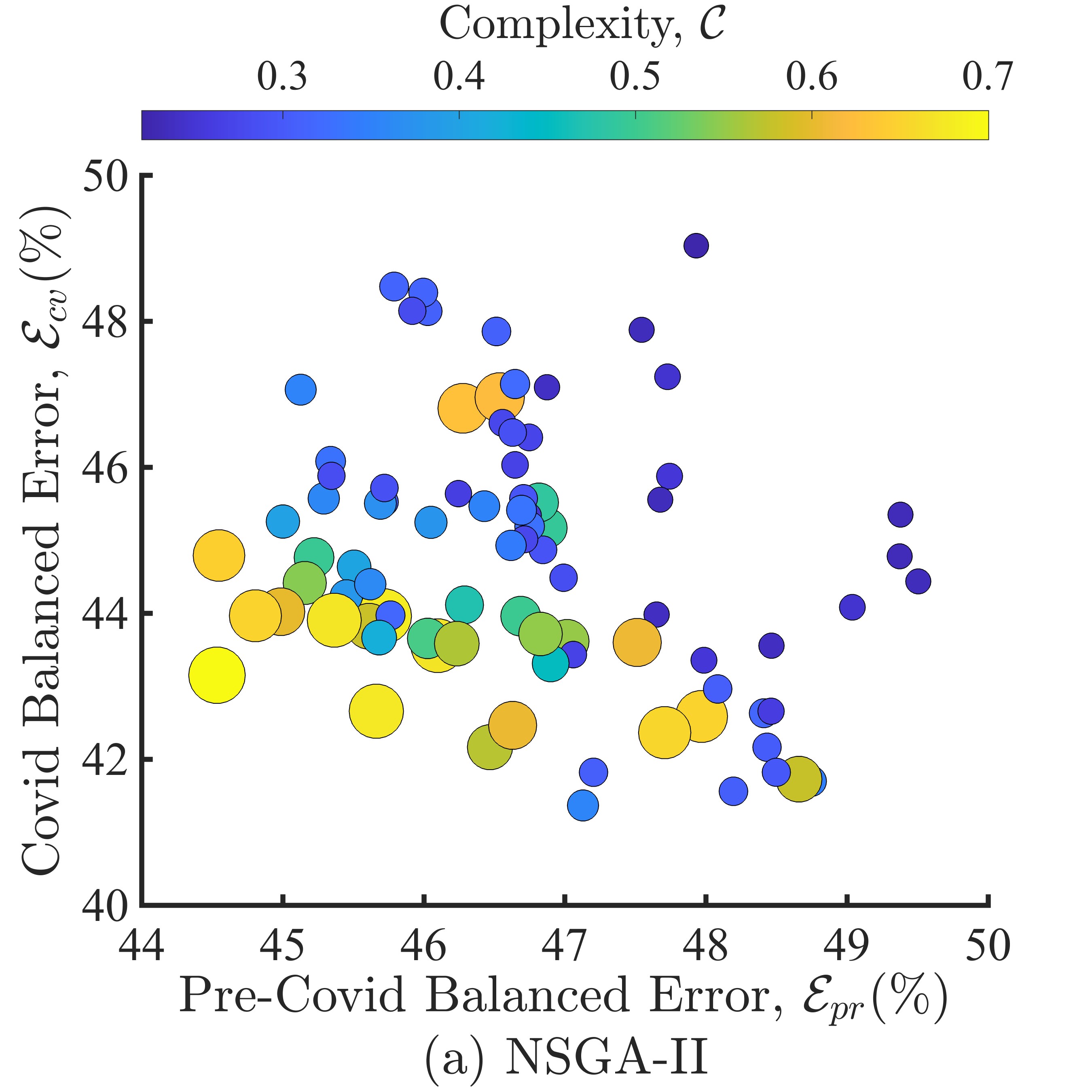}
\includegraphics[width=0.32\textwidth]{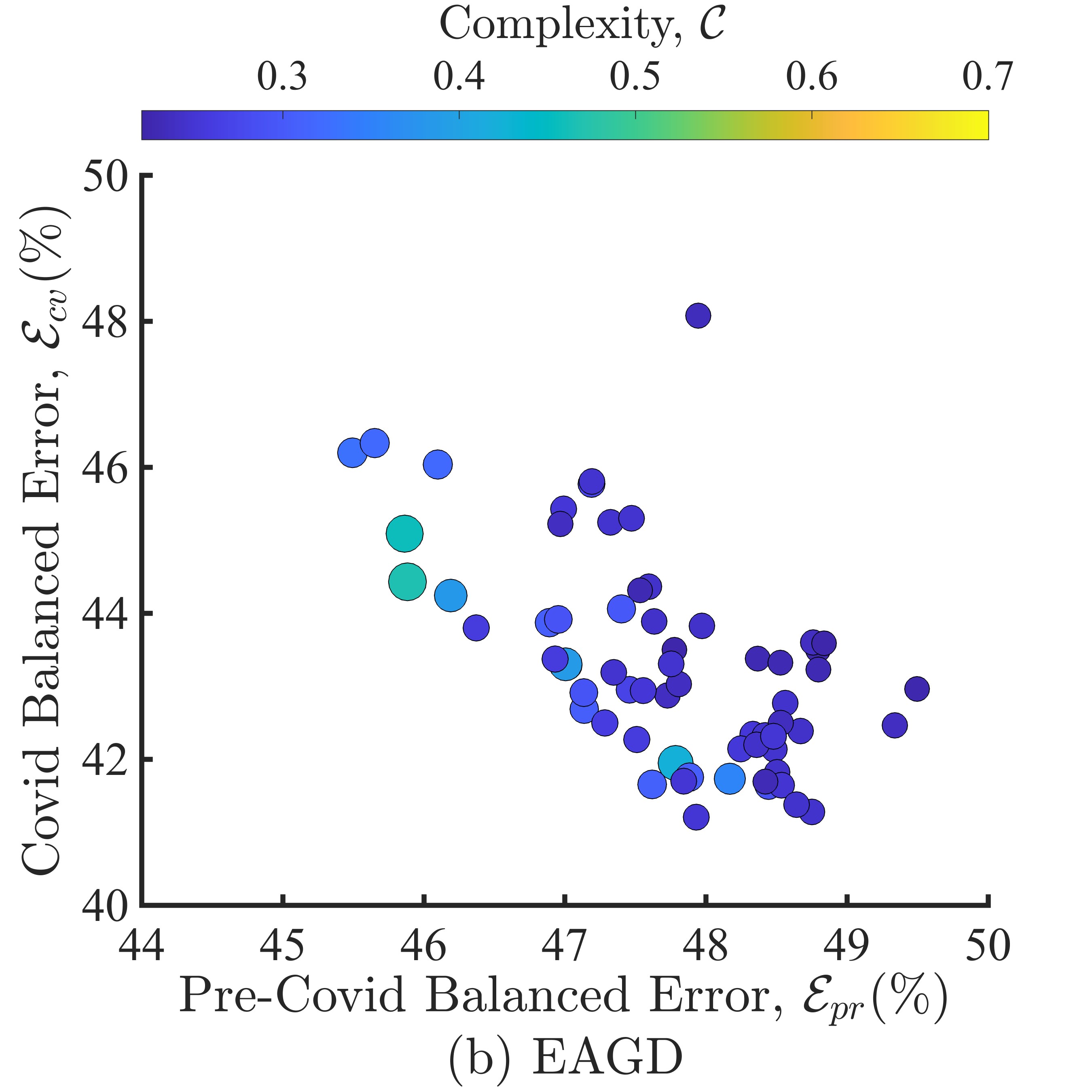} 

\caption{Trade-off in complexity ($\mathcal{C}$) and forecasting performances,\textit{ COVID balanced error} ($\mathcal{E}_{cv}$), and \textit{Pre-COVID balanced error} ($\mathcal{E}_{pr}$). Each circle represents a particular non-dominated architecture identified under the \textit{co-evolution} search scenario for the NASDAQ index. The values $\mathcal{C}$ are denoted by the radius and the color of circles; a larger radius and bright yellow color of a circle denotes a higher value of architectural complexity, whereas a smaller radius and dark blue color indicates the otherwise.}
\label{f:NSGAvsEAGD}
\end{SCfigure}

The search for non-dominated neural architectures involves navigating through a typically multi-model, deceptive, and noisy search space associated with the co-evolution problem~\cite{Yao:1999,Guyon:Isabelle:2003}. Accordingly, an effective search strategy is crucial to converge to the true Pareto front with diverse non-dominated neural architectures. To investigate the effects of different search strategies on the identified architectures, we compare the performance of NSGA-II and EAGD. Fig.~\ref{f:NSGAvsEAGD} shows the identified approximate Pareto fronts for the NASDAQ index. It is observed that the algorithms have been equally effective in reducing COVID balance error ($\mathcal{E}_{cv}$). Further, the results indicate a clear trade-off in complexity ($\mathcal{C}$) and pre-COVID balanced error ($\mathcal{E}_{pr}$); EAGD tends to select less complex architectures at the expense of higher values of $\mathcal{E}_{pr}$, see~Fig.~\ref{f:NSGAvsEAGD}(b). In contrast, the inclusion of non-geometric crossover~\cite{Ishibuchi:Tsukamoto:2010} in NSGA-II encourages the discovery of a relatively diverse set of architectures, as seen in Fig.~\ref{f:NSGAvsEAGD}(a). In particular, NSGA-II identified architectures with lower values of both $\mathcal{E}_{cv}$ and $\mathcal{E}_{pr}$; albeit with increased complexity as a trade-off. Given that the prediction performance is often preferred over the architectural complexity, we recommend NSGA-II with non-geometric crossover for the co-evolution problem.

\subsection{Co-evolution: Accommodating different decision  perspectives through a posteriori selection}
\label{subsec:results_Tournament_selection}

Most MOEAs, including NSGA-II and EAGD, are designed to obtain a diverse set of non-dominated solutions. Given that such non-dominated solutions are incomparable directly, the perspective of the DM is crucial to select the final solution through the \textit{a posteriori} selection from the identified Approximate Pareto Set, $\Gamma$. To this end, the combination of multiplicative preference relations~\cite{Zhang:Chen:2004} and the Multi-criteria Tournament Decision (MTD)~\citep{Parreiras:Vasconcelos:2009,Hafiz:Swain:MOEA:2020} is used, as discussed in Section~\ref{subsec:preferences}, and shown in Fig.~\ref{f:framework}. In particular, five distinct \emph{objective rankings} ($O$) are being considered to highlight distinct preference perspectives of the DM. Each ranking is defined as an \emph{order} of preference over \emph{COVID-era} performance ($O_{cv}$), \emph{architecture complexity} ($O_{\mathcal{C}}$) and \emph{pre-COVID-era} performance ($O_{pr}$), \textit{i.e.}, $O = \begin{bmatrix} O_{cv} & O_{\mathcal{C}} & O_{pr}\end{bmatrix}$. Further, the \emph{intensity} of preferences is set to strength `$9$' (\textit{i.e.}, $\mathcal{I}=9$) for all objective rankings. Both, the ranking scenarios ($O$) and the intensities ($\mathcal{I}$) at which such prioritization is supported, are then used to select an appropriate neural architecture from the approximated Pareto set, \textit{e.g.}~Section~\ref{subsec:preferences}.

\begin{table*}[!t]
  \centering
  \footnotesize
   \caption{Effects of preferences on \textit{ a posteriori} architecture selection for NASDAQ: NSGA-II}
  \label{t:nasdaq_Pareto_perspectives}%
  \begin{adjustbox}{width=0.8\textwidth}
  \begin{threeparttable}
    \begin{tabular}{ccccccccc}
    \toprule
    \makecell{$^\dagger$\textbf{Rankings}\\\textbf{\& Selected}\\\textbf{Architecture}} & \makecell{\textbf{Selected}\\ \textbf{Features}\\ $|X|$} & \makecell{\textbf{Hidden}\\\textbf{Layer-1}\\ $(s^1,f^1)$} & \makecell{\textbf{Hidden}\\\textbf{Layer-2}\\ $(s^2,f^2)$} & \makecell{\textbf{Complexity,}\\$\mathcal{C}$} & \makecell{$^{\dagger\dagger}$\textbf{COVID}\\ \textbf{Data,}\\$\mathcal{D}_{test}$} & \makecell{$^{\dagger\dagger}$\textbf{COVID}\\ \textbf{Data,}\\$\mathcal{D}_{hold}$} & \makecell{$^{\dagger\dagger}$\textbf{Pre-COVID}\\ \textbf{Data,}\\$\mathcal{D}_{pr}$} \\[2ex]
    \midrule

    \multirow{2}[2]{*}{\makecell{$O_1$, $\mathcal{A}_1$}} & \multirow{2}[1]{*}{11} & 18  & - & \multirow{2}[1]{*}{0.27} & 61.84 & 55.93 & 55.47 \\
    &       & \textit{tansig} & - &       & (0.16)  & (0.13)  & (0.06) \\ [1ex]
    \midrule

    \multirow{2}[2]{*}{\makecell{$O_2, \mathcal{A}_2$}} & \multirow{2}[0]{*}{11} & 32 & -    & \multirow{2}[0]{*}{0.30} & 63.17 & 56.18 & 56.34 \\
    &       & \textit{tansig} & - &   & (0.20)  & (0.14)  & (0.06) \\ [1ex]
    \midrule

    \multirow{2}[2]{*}{\makecell{$O_3, \mathcal{A}_3$}} & \multirow{2}[0]{*}{10} & 35 & 32 & \multirow{2}[0]{*}{0.47} & 60 & 56.26 & 56.41 \\
     &       & \textit{tansig} & \textit{logsig} &       & (0.13)  & (0.13)  & (0.08)\\ [1ex]
    \midrule

    \multirow{2}[2]{*}{\makecell{$O_4, \mathcal{A}_4$}} & \multirow{2}[1]{*}{14} & 123   & 64    & \multirow{2}[1]{*}{0.65} & 58.76 & 56.09 & 58.33 \\
    &       & \textit{tansig} & \textit{logsig} &  & (0.11)  & (0.13)  & (0.12)\\[1ex]

    \midrule
    \multirow{2}[2]{*}{\makecell{$O_5, \mathcal{A}_5$}} & \multirow{2}[1]{*}{13} & 48   & -    & \multirow{2}[1]{*}{0.36} & 63.32 & 56.35 & 56.36 \\
    &       & \textit{tansig} & - &  & (0.21)  & (0.14)  & (0.07)\\[1ex]
		
    \bottomrule
    \end{tabular}
    \begin{tablenotes}
      \footnotesize
      \item $^\dagger$ - ranking denotes DM's preference towards search objectives, \textit{i.e.}, \textit{COVID performance}, \textit{complexity} and \textit{Pre-COVID performance} and it is denoted by $O = \begin{bmatrix} O_{cv} & O_{\mathcal{C}} & O_{pr}\end{bmatrix}$; $O_1, O_2, \dots O_5$ denote different objective rankings and are given as follows: $O_1 = \begin{bmatrix} 1 & 1 & 1 \end{bmatrix}$, $O_2 = \begin{bmatrix} 1 & 2 & 3 \end{bmatrix}$, $O_3 = \begin{bmatrix} 1 & 2 & 1 \end{bmatrix}$, $O_4 = \begin{bmatrix} 2 & 3 & 1 \end{bmatrix}$, and $O_5 = \begin{bmatrix} 1 & 3 & 3 \end{bmatrix}$
      \item $^{\dagger\dagger}$ - forecasting performance in terms of overall accuracy/hit-rate (in percentage) and Matthews Correlation Coefficient (MCC); a higher value is desirable for both metrics. The values of MCC are shown inside parentheses.
    \end{tablenotes}
  \end{threeparttable}
 \end{adjustbox}
\end{table*}%

Table~\ref{t:nasdaq_Pareto_perspectives} lists the objective rankings being considered along with the corresponding selected non-dominated architectures from the Pareto set identified by NSGA-II for the NASDAQ index. It provides the details about the selected neural topologies, their complexity ($\mathcal{C}$), as well as classification performances in COVID and Pre-COVID time periods. The classification performance is measured in terms of overall accuracy as well as using MCC (see Section~\ref{subsec:metrics}). Given that the forecasting model is essentially a binary classifier, the overall accuracy is equivalent to the well-known financial prediction metric `\emph{Hit-Rate}'. 
In the following, the result corresponding to each decision perspective is discussed individually for ease of interpretation:
\begin{itemize} [leftmargin=*]
    \item \textit{Perspective-1: Indifference}, $O_{cv}\!\thicksim\!O_\mathcal{C}\!\thicksim\!O_{pr}$: In the first decision perspective, the DM is \emph{indifferent}, \textit{i.e.}, all objectives have equal preference. This is reflected by $O_1 = \begin{bmatrix} 1 & 1 & 1 \end{bmatrix}$ in Table~\ref{t:nasdaq_Pareto_perspectives}. A comparatively sparse neural architecture, $\mathcal{A}_1$, is selected in this scenario; $11$ features and a single hidden layer with $18$ neurons. A relatively lower pre-COVID classification performance (MCC $= 0.06$) is obtained with this architecture. This scenario serves as a \textit{baseline} for the comparison with the other perspectives. 
    
    \item \textit{Perspective-2}, $O_{cv}\!\succ\!O_\mathcal{C}\!\succ\!O_{pr}$: The DM emphasizes \emph{COVID} performance and secondly complexity, which is reflected by $O_2 = \begin{bmatrix} 1 & 2 & 3 \end{bmatrix}$. The effects of these preferences are clearly visible in the selected architecture, $\mathcal{A}_2$, which is relatively more complex than $\mathcal{A}_1$, \textit{i.e.}, $\mathcal{C}(A_2)>\mathcal{C}(\mathcal{A}_1)$, albeit with comparatively better \emph{COVID} classification performance. 
    
    \item \textit{Perspective-3},
    $O_{cv}\!\thicksim\!O_{pr}\succ\!O_\mathcal{C}$:  Here the DM prefers the classification performance in both time-periods over complexity, \textit{i.e.}, $O_3 = \begin{bmatrix} 1 & 2 & 1 \end{bmatrix}$. While the selected architecture $\mathcal{A}_3$ is more complex than the baseline $\mathcal{A}_1$, the corresponding classification performance is better than $\mathcal{A}_1$ in both \emph{pre-} and \emph{within-COVID} periods. Note that $\mathcal{A}_3$ is \textit{semi-heterogeneous} with different activation functions in each hidden layer.  
    
    \item \textit{Perspective-4}, $O_{pr}\!\succ\! O_{cv}\!\succ\!O_\mathcal{C}$: The DM emphasizes most on the performance over the \emph{pre-COVID} period and the least on complexity, \textit{i.e.}, $O_4 = \begin{bmatrix} 2 & 3 & 1\end{bmatrix}$. As expected, the selected architecture in this scenario, $\mathcal{A}_4$, improves \emph{pre-COVID} performance at the expense of higher complexity; $\mathcal{A}_4$ is the most complex architecture among all scenarios with two dense hidden layers, see Table~\ref{t:nasdaq_Pareto_perspectives}. $\mathcal{A}_4$ is also \textit{semi-heterogeneous}. 
    
    \item \textit{Perspective-5}, $O_{cv}\!\succ\!O_\mathcal{C}\!\thicksim\!O_{pr}$: In the final perspective, the DM prefers a better \emph{within-COVID} performance, whereas they are indifferent to the complexity and \emph{pre-COVID} performance, \textit{i.e.}, $O_5 = \begin{bmatrix} 1 & 3 & 3\end{bmatrix}$. The architecture $\mathcal{A}_5$ has better classification performances in \emph{pre-} and \emph{within-COVID} periods with a trade-off in complexity compared to $\mathcal{A}_1$.
    
\end{itemize}

The comparative evaluation of these scenarios further underlines a trade-off in \emph{architecture complexity} and \emph{pre-COVID} performance in the identified architectures. Nevertheless, all selected architectures ($\mathcal{A}_1, \mathcal{A}_2, \dots, \mathcal{A}_5$, Table~\ref{t:nasdaq_Pareto_perspectives}) provide similar performance on \emph{out-of-sample} COVID dataset ($\mathcal{D}_{hold}$), \textit{i.e.}, overall classification accuracy (hit rate) $\approx 56\%$ and MCC$\approx0.14$.

\begin{table*}[!t]
  \centering
  \caption{Results of baseline neural design: \textit{a priori} feature selection with \textit{rules of thumb} for neural topology selection}
  \label{t:resrot}%
  \begin{adjustbox}{max width=0.7\textwidth}
  \scriptsize
  \begin{threeparttable}
    \begin{tabular}{cccccccc}
    \toprule
    \multirow{2}[2]{*}{\textbf{Scenario}} & \multirow{2}[4]{*}{\makecell{$^\ddagger$\textbf{Topology}\\ \textbf{Rule}}} & \multirow{2}[4]{*}{\makecell{\textbf{Layer}\\\textbf{Size}\\$(s^1)$}} & \multirow{2}[4]{*}{\makecell{\textbf{Layer}\\\textbf{Size}\\$(s^2)$}} & \multirow{2}[4]{*}{\makecell{\textbf{Complexity}\\$\mathcal{C(\cdot)}$}} & \multicolumn{3}{c}{$^{\dagger\dagger}$\textbf{Classification Performance}} \\
    \cmidrule{6-8}          &       &       &       &       & \makecell{ \textbf{COVID}\\$\mathcal{D}_{test}$} & \makecell{ \textbf{COVID}\\$\mathcal{D}_{hold}$} & \makecell{ \textbf{Pre-COVID}\\$\mathcal{D}_{pr}$} \\
    \midrule
    \multirow{6}[4]{*}{\makecell{PCA + RoT\\44 principals\\ (99.99\% of\\ Variance\\ Explained)}} & Kolmogorov & 89    & - & 0.3597 & \makecell{55.90\\ (0.02)} & \makecell{52.35\\ (0.05)} & \makecell{53.44\\ (\ul{0})} \\[1ex]	 		
     \cmidrule{2-8} & Hush & 48    & -     & 0.3995 & \makecell{55.61\\ (\ul{0})} & \makecell{51.63 \\ (0.03)} & \makecell{54.16\\ (\ul{0})} \\[1ex]			
     \cmidrule{2-8} & Wang & 29    & - & 0.3323 & \makecell{56.87\\ (0.04)} & \makecell{51.56\\ (0.03)} & \makecell{53.85\\ (\ul{0})}\\[1ex]			
     \cmidrule{2-8} & Ripley & 23    & - & 0.3297 & \makecell{56.37\\ (0.01)} & \makecell{51.74\\ (0.03)} & \makecell{54.30\\ (\ul{0})} \\[1ex]			
     \cmidrule{2-8} & Fletcher-Goss & 15    & -     & 0.3271 & \makecell{57.53\\ (0.05)} & \makecell{51.12\\ (0.02)} & \makecell{54.55\\ (\ul{0})} \\[1ex] 			
    \cmidrule{2-8} & Huang & 57    & 19    & 0.5101 & \makecell{57.36\\ (0.01)} & \makecell{51.93\\ (0.03)} & \makecell{54.38\\ (\ul{0})} \\

    \midrule
     \multirow{12}{*}{\makecell{mRmR + RoT\\17 features}} & \multirow{2}[2]{*}{Kolmogorov} & \multirow{2}{*}{35}    & \multirow{2}{*}{-}     & \multirow{2}{*}{0.3419} & 57.48 & 51.10 & 54.57 \\
           &       &     &     &       & (0.01)  & (0.02)  & (0.02) \\
    \cmidrule{2-8} & \multirow{2}[2]{*}{Hush} & \multirow{2}{*}{68} & \multirow{2}{*}{-} & \multirow{2}{*}{0.4285} & 57.53 & 50.44 & 54.22 \\
           &       &      &      &       & (0.01)  & (\ul{0})  & (0.01) \\
    \cmidrule{2-8} & \multirow{2}[2]{*}{Wang} & \multirow{2}{*}{11} & \multirow{2}{*}{-} & \multirow{2}{*}{0.2789} & 56.95 & 51.41 & 54.95 \\
             &    &     &      &       & (0.01)  & (0.02)  & (0.01) \\
    \cmidrule{2-8} & \multirow{2}[1]{*}{\makecell{Ripley \&\\Fletcher-Goss}} & \multirow{2}{*}{10}  & \multirow{2}{*}{-} & \multirow{2}{*}{0.2762} & 56.06 & 51.54 & 54.90 \\
          &        &      &      &       & (\ul{0})  & (0.03)  & (0.02) \\
    \cmidrule{2-8} & \multirow{2}[2]{*}{Huang} & \multirow{2}{*}{57} & \multirow{2}{*}{19} & 0.5164 & 55.24 & 50.51 & 55.35 \\
          &       &     &      &       & (\ul{0})  & (\ul{0})  & (0.02) \\

    \midrule
    \multirow{12}{*}{\makecell{CFS + RoT\\29 features}} & \multirow{2}{*}{Kolmogorov} & \multirow{2}{*}{59} & \multirow{2}{*}{-} & \multirow{2}{*}{0.4637} & 56.72 & 50.70 & 53.27 \\
          &       &       &       &       & (\ul{-0.04}) & (\ul{0})  & (0.01) \\
    \cmidrule{2-8}          & \multirow{2}{*}{Hush} & \multirow{2}{*}{116} & \multirow{2}{*}{-} & \multirow{2}{*}{0.6133} & 57.18 & 51.03 & 52.29 \\
          &       &       &       &       & (\ul{-0.03}) & (0.02)  & (\ul{0}) \\
    \cmidrule{2-8}          & \multirow{2}{*}{Wang} & \multirow{2}{*}{19} & \multirow{2}{*}{-} & \multirow{2}{*}{0.3587} & 55.71 & 51.01 & 53.96 \\
          &       &       &       &       & (\ul{-0.02}) & (0.01)  & (0.02) \\
    \cmidrule{2-8}          & \multirow{2}{*}{Ripley} & \multirow{2}{*}{16} & \multirow{2}{*}{-} & \multirow{2}{*}{0.3508} & 56.16 & 51.71 & 53.84 \\
          &       &       &       &       & (\ul{-0.02}) & (0.04)  & (0.02) \\
    \cmidrule{2-8}          & \multirow{2}{*}{Fletcher-Goss} & \multirow{2}{*}{13} & \multirow{2}{*}{-} & \multirow{2}{*}{0.3429} & 55.53 & 52.09 & 53.70 \\
          &       &       &       &       & (\ul{-0.04}) & (0.05)  & (0.02) \\
    \cmidrule{2-8} & \multirow{2}{*}{Huang} & \multirow{2}{*}{57} & \multirow{2}{*}{19} & \multirow{2}{*}{0.5752} & 55.98 & 50.81 & 53.36 \\
          &       &       &       &       & (\ul{-0.01}) & (0.02)  & (0.01) \\

    \bottomrule
    \end{tabular}%
    \begin{tablenotes}
      \scriptsize
      \item $^\ddagger$ See~\ref{sec:app} for the details about rules of thumb; $^\dagger$ `$tanh$' is selected as the activation function for hidden layer/s; $^{\dagger\dagger}$ forecasting performance in terms of overall classification accuracy/hit-rate (in percentage) and Matthews Correlation Coefficient (MCC). The values of MCC are shown inside parentheses. The \emph{underlined} MCC values indicate \emph{biased} prediction behavior towards index \emph{rise}.
    \end{tablenotes}
  \end{threeparttable}
 \end{adjustbox}
\end{table*}%

\subsection{Comparative Evaluation with Baseline Approaches}
\label{subsec:compEval}

The details of neural forecasting models for the NASDAQ index, which are designed following the first baseline approach, are shown in Table~\ref{t:resrot}. In particular, the input dimensionality of the neural network models is reduced first via either PCA or \emph{filter-}based feature selection (mRmR or CFS), as discussed in Section~\ref{sec:baseline1}. Next, the neural topologies are determined using various \emph{rules-of-thumb} in~\ref{sec:app}. The results show that most neural architectures designed using this baseline have very low values of MCC ($<0.01$, highlighted by underlining in Table~\ref{t:resrot}). This behavior can be explained by the tendency of neural networks to be a `\textit{brute}' classifier, \textit{i.e.}, \textit{predict only index rises}. Similar behavior has also been identified in the earlier investigations (\textit{e.g.}, \textit{see}~\cite{kumar:Meghwani:2016}), and is highly undesirable. Note that CFS is not considered for further analysis as it led to `brute' classifiers with all rules-of-thumb; see MCC values with CFS for $\mathcal{D}_{test}$ in Table~\ref{t:resrot}.

\begin{figure*}[!t]
   \centering
   \includegraphics[width=0.85\textwidth]{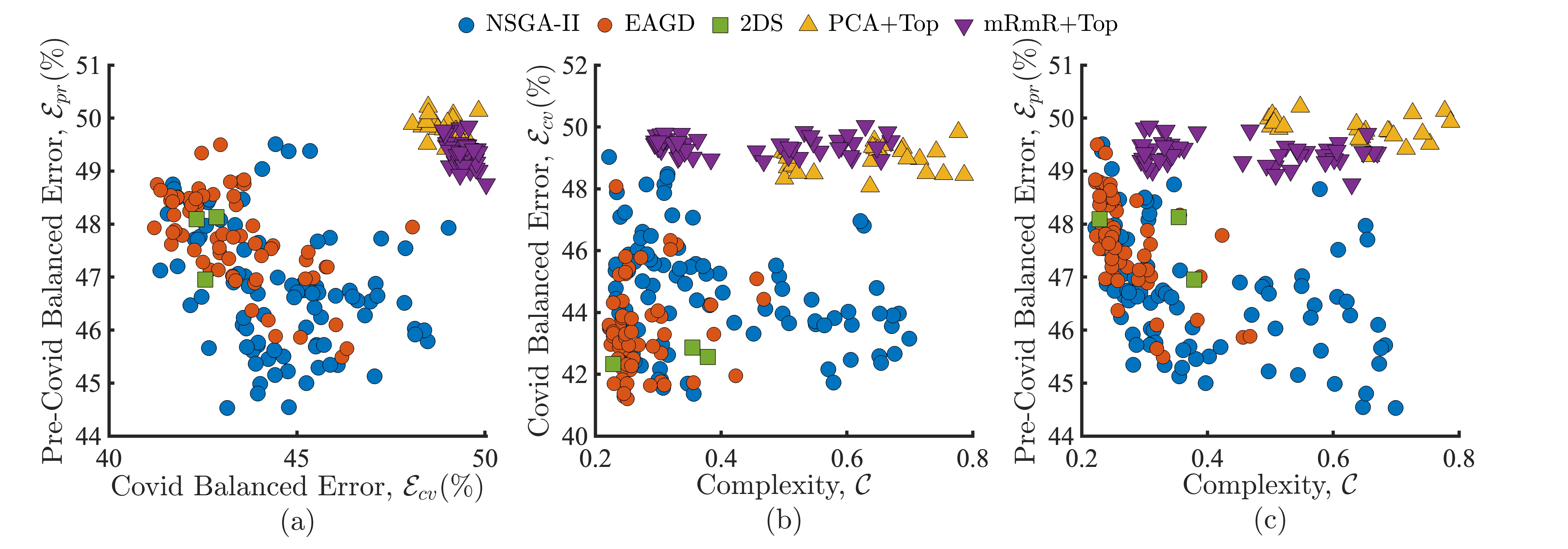} 
   \caption{Comparative evaluation of neural design approaches. Legends: NSGA-II (\emph{co-evolution}); EAGD (\emph{co-evolution}); 2DS (\emph{scalerized co-evolution}); PCA+Top (\emph{PCA and Topology Optimization}); PCA+RoT (\emph{PCA and Rule of Thumbs}); mRmR+Top (\emph{mRmR and Topology Optimization}); mRmR+RoT (\emph{mRmR and Rule of Thumbs}); CFS+RoT (\emph{CFS and Rule of Thumbs}).}
   \label{f:APF_MOEA}
\end{figure*}

Next step of the comparative evaluation focuses on the performance of the second (Section~\ref{sec:baseline2}) and third (Section~\ref{subsec:baselineScalerized}) baselines on the NASDAQ index. The second baseline optimizes neural topologies after the dimensionality reduction using either PCA or mRmR. In contrast, the third baseline is essentially a \textit{scalarized} version of the proposed multi-objective co-evolution. Note that the second baseline, only topology optimization, identified a large number of neural topologies, so we omit the detailed description of neural architectures (similar to Table~\ref{t:resrot}) for sake of brevity. Instead, we compare the performance of the identified neural models following these baselines using the two-dimensional Pareto front plots shown in Fig.~\ref{f:APF_MOEA}. Note that the \textit{scalarized} approach requires the \textit{a priori} specification of the preference weights, $[\theta_{\rm{E}}, \theta_\mathcal{C}]$, from the DM. The results obtained using the following three distinct preference scenarios are shown in Fig.~\ref{f:APF_MOEA} to highlight the different preferences of the DM: (1) \emph{Efficacy} over \emph{Complexity}: $[\theta_{\rm{E}}, \ \theta_\mathcal{C}] = [0.75, \ 0.25]$, (2) \emph{Balanced} Scenario: $[\theta_{\rm{E}}, \ \theta_\mathcal{C}] = [0.50, \ 0.50]$, and (3) \emph{Complexity} over \emph{Efficacy}: $[\theta_{\rm{E}}, \ \theta_\mathcal{C}] = [0.25, \ 0.75]$.

\begin{figure}[!t]
\centering
\includegraphics[width=0.4\textwidth]{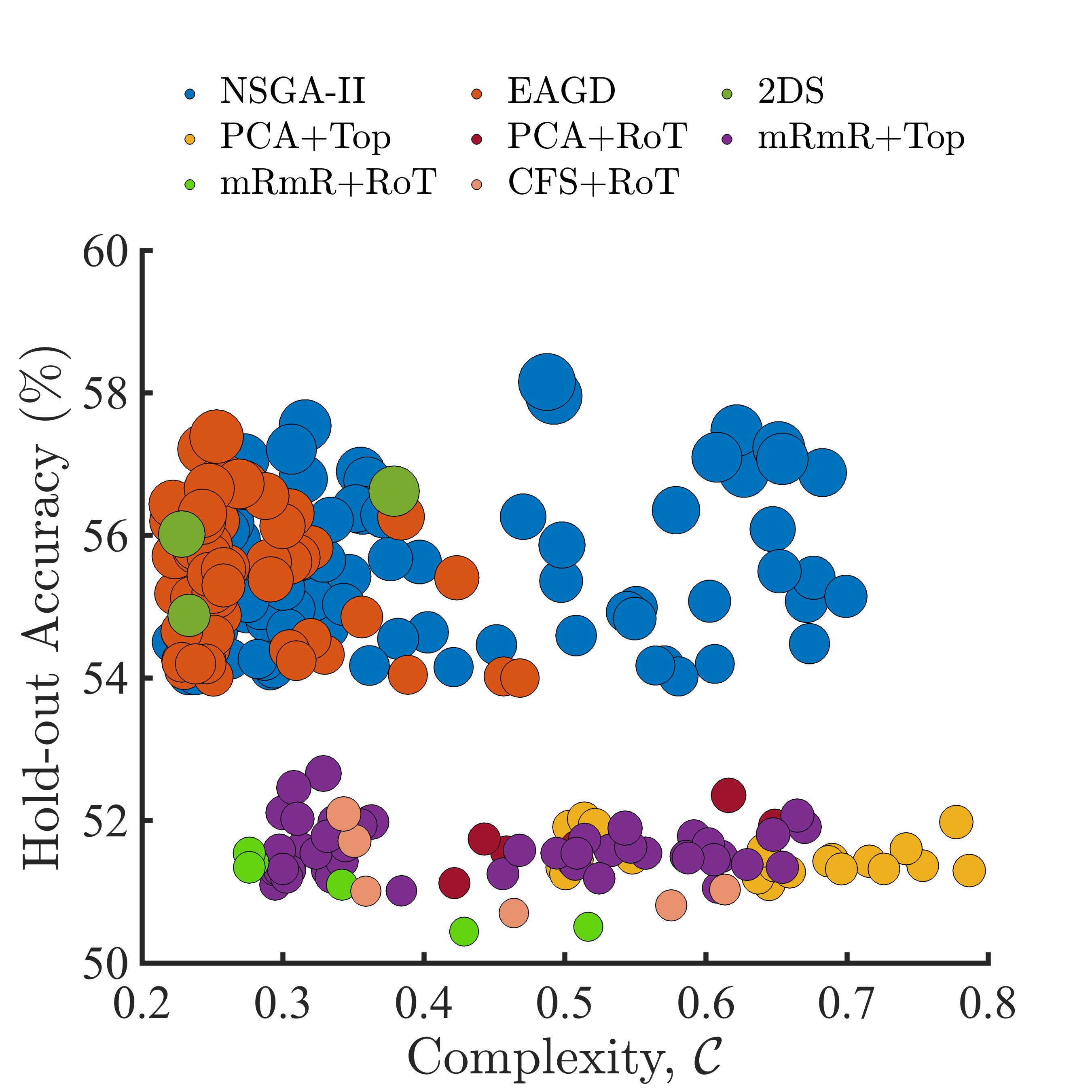} 
\caption{Comparative evaluation of neural design approaches on the \emph{out-of-sample} data, $\mathcal{D}_{hold}$. Each circle represents a neural architecture for the NASDAQ index. The MCC values associated with each architecture are reflected by the circle size, \textit{i.e.}, a larger circle denotes a higher value of MCC. Legends: NSGA-II (\emph{co-evolution}); EAGD (\emph{co-evolution}); 2DS (\emph{scalerized co-evolution}); PCA+Top (\emph{PCA and Topology Optimization}); PCA+RoT (\emph{PCA and Rule of Thumbs}); mRmR+Top (\emph{mRmR and Topology Optimization}); mRmR+RoT (\emph{mRmR and Rule of Thumbs}); CFS+RoT (\emph{CFS and Rule of Thumbs}).}
\label{f:comp_holdout}
\end{figure}

The results in Fig.~\ref{f:APF_MOEA} indicate that most of the architectures identified by the second baseline (optimization of only topologies) are dominated by the \textit{proposed} and \textit{scalarized} approach. It is, therefore, safe to infer that the co-evolution of the feature subset and the neural topology can yield comparatively better results. Further, it is interesting to see that the architectures identified following the \textit{scalarized} approach are comparable to the proposed approach and lie in the knee-point region of the Pareto front when only the COVID error ($\mathcal{E}_{cv}$) and the complexity ($\mathcal{C}$) are considered, see Fig.~\ref{f:APF_MOEA}(b). However, the \textit{scalarized} architectures tend to have higher pre-COVID errors ($\mathcal{E}_{pr}$) than the \textit{proposed} approach, as seen in Fig.~\ref{f:APF_MOEA}(a) and (c). This is expected and can be explained as follows: the \textit{scalarized} approach was designed to attain only a pre-specified threshold for $\mathcal{E}_{pr}$, as discussed earlier. In contrast, the \textit{proposed} approach can better balance performance over both COVID and pre-COVID data owing to the latter's inclusion in the multi-objective formulation.

For sake of convenience, overall classification accuracy (\emph{hit rate}) and MCC on the \emph{out-of-sample} dataset ($\mathcal{D}_{hold}$) of the neural models designed using all comparative baselines and the proposed co-evolution approach are depicted in Fig.~\ref{f:comp_holdout}. It is clear that both co-evolution approaches, scalerized (baseline-3) and proposed multi-objective formulation, outperform baselines-1 and 2. The significant improvement in MCC with the co-evolution approaches is especially noteworthy. The limitations of decoupled neural design approaches (baseline-1 and 2) can primarily be attributed to the input dimensionality reduction approaches (PCA, mRmR, CFS), which are often unsuccessful in accounting for nonlinear feature interactions.

\begin{table*}[!t]
  \centering
  \footnotesize
   \caption{Outcomes of Hommel’s post-hoc procedure on out-of-sample dataset ($\mathcal{D}_{hold}$)}
  \label{t:pval}%
  \begin{adjustbox}{width=0.85\textwidth}
  \begin{threeparttable}
    \begin{tabular}{lcccc|ccc}
    \toprule
    \multirow{2}[2]{*}{\textbf{Scenario}} & \multirow{2}[2]{*}{\textbf{Approach}} & \multicolumn{3}{c}{\textbf{Accuracy}} & \multicolumn{3}{c}{\textbf{MCC}} \\
    \cmidrule{3-8}   &       & \textbf{Mean $\pm$ SD} & $p-$\textbf{value} & \makecell{ $^\ddagger$\textbf{Adjusted}\\$p-$\textbf{value}} & \textbf{Mean $\pm$ SD} & $p-$\textbf{value} & \makecell{$^\ddagger$\textbf{Adjusted}\\$p-$\textbf{value}} \\
    \midrule
    \multirow{2}{*}{Na\"ive} & Random & 49.38 $\pm$ 4.6 & 1.0E-18 & 6.3E-03$^{\ast\ast}$ & -0.01 $\pm$ 0.09 & 5.7E-20 & 6.3E-03$^{\ast\ast}$ \\[0.1ex]
    & Brute & 49.45 & 3.8E-22 & 5.6E-03$^{\ast\ast}$ &  0 & 4.4E-22 & 5.6E-03$^{\ast\ast}$ \\[0.1ex]
    \midrule
    \multirow{3}{*}{Baseline-1} & \makecell{PCA + Kolmogorov} & 52.35 $\pm$ 2.9  & 2.5E-10 & 1.3E-02$^{\ast\ast}$ & 0.05 $\pm$ 0.07 & 4.6E-09 & 1.7E-02$^{\ast\ast}$ \\[0.1ex]
    & \makecell{mRmR + Fletcher-Goss} & 51.54 $\pm$ 2.9 & 2.3E-13 & 7.1E-03$^{\ast\ast}$ & 0.03 $\pm$ 0.08 & 9.5E-14 & 7.1E-03$^{\ast\ast}$ \\[0.1ex]
    & \makecell{CFS + Fletcher-Goss} & 52.09 $\pm$ 2.6 & 3.1E-10 & 1.7E-02$^{\ast\ast}$ & 0.05 $\pm$ 0.07 & 4.3E-10 & 1.3E-02$^{\ast\ast}$ \\[0.1ex]
    \midrule
    \multirow{2}{*}{Baseline-2} & \makecell{PCA + Top} & 52.20 $\pm$ 2.4 & 1.6E-10 & 1.0E-02$^{\ast\ast}$ & 0.05 $\pm$ 0.06 & 1.3E-10 & 8.3E-03$^{\ast\ast}$ \\[0.1ex]
    & \makecell{mRmR + Top} & 52.18 $\pm$ 3.4 & 3.5E-11 & 8.3E-03$^{\ast\ast}$ & 0.05 $\pm$ 0.09 & 4.3E-10 & 1.0E-02$^{\ast\ast}$ \\[0.1ex]
    \midrule
    Baseline-3 & \makecell{Scalerized\\(2DS)} & 56.62 $\pm$ 3.1 & 4.6E-01 & 2.5E-02$^{\ast\ast}$ & 0.13 $\pm$ 0.08 & 3.1E-01 & 2.5E-02$^{\ast\ast}$ \\[0.1ex]
    \midrule
    \multirow{2}{*}{Co-evolution} & \makecell{EAGD}  &  57.93 $\pm$ 2.6 & 8.9E-01 & 5.0E-02 & 0.19 $\pm$ 0.07 & 8.8E-01 & 5.0E-02\\[0.1ex]
    & \makecell{NSGA-II} & 58.07 $\pm$ 3.0 & -     & - &  0.19 $\pm$ 0.07 & -     & - \\[0.1ex]
    \bottomrule
    \end{tabular}%
    \begin{tablenotes}
      \footnotesize
      \item  $^\ddagger$  null hypothesis states that a particular approach is better than the Co-evolution (NSGA-II); Hommel's posthoc procedure suggests that all null-hypotheses ($H_0$) with APV$\leq0.025$ should be rejected at $\alpha=0.05$ level (95\% confidence interval); $^{\ast\ast}$ denotes that null-hypothesis is rejected
    \end{tablenotes}
  \end{threeparttable}
 \end{adjustbox}
\end{table*}%

The statistical significance of the results is determined through non-parametric statistical tests, following the guidelines in~\cite{Garcia:Salvador:2010}. To this end, the neural architecture with the best performance on $\mathcal{D}_{hold}$ is selected first for each baseline as well as the proposed co-evolution approach. Table~\ref{t:pval} shows the classification performance of these architectures, both in terms of overall accuracy and MCC. Note that for this part of the analysis, two na\"ive classifiers are also being considered: \emph{na\"ive random} which randomly predicts \emph{up} or \emph{down} label for each test instance; and \emph{na\"ive brute} which always predicts \emph{up} movement irrespective of the test instance.

The results of Friedmann's two-way analysis by ranks support the rejection of the null-hypothesis that all the architectures have equal performance with the following $p-$values:  1.08E-10 (overall accuracy),  1.12E-10 (MCC). Further, the top three neural architectures identified as per Friedman rankings are from the following approaches: Co-evolution (NSGA-II), Co-evolution (EAGD), and Scalerized (2DS). Based on these rankings, Co-evolution (NSGA-II) serves as the \emph{control} approach in the subsequent post-hoc analysis, in which the control approach is compared with the remaining approaches using a set of null hypotheses. Each hypothesis states that the approach being compared identifies significantly better architecture than co-evolution (NSGA-II). The adjusted $p-$values (APV) for this test are determined using Hommel's post-hoc analysis~\cite{Garcia:Salvador:2010}. The results of this analysis clearly show that all \textit{null-hypotheses}, except with co-evolution (EAGD), can safely be rejected at the significance level $\alpha=0.05$, see Table~\ref{t:pval}.

\section{Discussion \& Conclusions}
\label{sec:conclusions}

The development of forecasting models is a critical supporting tool that helps the decision-maker minimize the risk arising from intrinsic and COVID-19 induced uncertainties in stock markets. Neural networks are among the most performant forecasting models for this task, provided their architecture is carefully designed. The design of neural architecture is a multi-criteria problem that requires a careful balance of \emph{parsimony} (\textit{selection of features and topology}) with \emph{efficacy} (\textit{forecasting performance over pre- and within-COVID data}). This study proposed a new co-evolution approach that simultaneously targets feature selection, topological complexity, and multi-dataset learning to address these issues. In addition, a search framework consisting of diversity-focused MOEAs and an \textit{a posteriori} architecture selection was proposed. It was shown that the introduction of the non-geometric recombination operator allows for the identification of a set of diverse neural architectures in terms of complexity as well as the forecasting performance over multiple datasets. This gives the Decision Maker (DM) a higher degree of selection freedom for a preferred criterion. Further, a combination of multiplicative preference relations and a tournament decision was used to select architecture \textit{a posteriori} from the identified Approximate Pareto Set (APS). It was shown that this combination could embed the preferences of the DM into the final neural architecture selection. 

The efficacy of the proposed co-evolution was demonstrated by considering a total of $21$ different neural architecture design approaches, which were broadly categorized into three comparative \emph{neural design baselines}. These comparative baselines represent neural forecasting models in most existing investigations, as discussed in Section~\ref{sec:compeval}. The comparative evaluation demonstrated a statistically significant improvement with the co-evolution. 

Further, this investigation provided yet another empirical result supporting market inefficiency. However, more importantly, the results of comparative evaluations could, in part, explain the disparity in conclusions about the forecasting performance of shallow neural networks in the existing research. In particular, the results show that it is possible to obtain insufficient forecasting models which conform to market efficiency or are \emph{brute} in nature 
with an improper selection of features and neural topology. 
The proposed \emph{co-evolution} approach can overcome such problems by identifying optimal combinations of features and neural topology, which results in balanced forecasting models.

This investigation focused on the forecasting of stock indices, the co-evolution framework proposed here is generic and can be used for any dataset with a moderate number of features. Further, 
the forecasting performance was evaluated only in terms of classification metrics; the construction trading strategies using the model prediction is left as a future exercise. However, the co-evolution of neural architecture can further be improved by considering other metrics which are closer to profitability, \textit{e.g.}, see \emph{mean profit rate} in~\cite{Liu:Wang:2019}.

\appendix

\section{Empirical Rules for Neural Design}
\label{sec:app}

Kolmogorov's Theorem: $(s^1, \ s^2) \rightarrow \Big( 2 n_f + 1, \ 0 \Big)$; Hush's Rule: $(s^1, \ s^2) \rightarrow \Big( \lceil c_1 \times n_f \rceil, \ \lceil c_2 \times m \rceil \Big)$, where, $c_1 \in [2,4]$, $c_2 \in [2,3]$; Wang's Rule : $(s^1, \ s^2) \rightarrow \Big( \bigg \lceil \displaystyle\frac{2 \times n_f}{3} \bigg \rceil, \ 0\Big)$; Ripley's Rule : $(s^1, \ s^2) \rightarrow \Big( \bigg \lceil \dfrac{n_f + m}{2} \bigg \rceil, \ 0\Big)$; Fletcher-Goss's Rule : $(s^1, \ s^2) \rightarrow \Big( \lceil 2\sqrt{n_f}+m \rceil , \ 0 \Big)$; Huang's Rule : $(s^1, \ s^2) \rightarrow \Bigg( \bigg \lceil \sqrt{(m+2)N} + \big(2\sqrt{\frac{N}{m+2}} \big) \bigg \rceil, \ \bigg \lceil m\sqrt{\frac{N}{m+2}} \bigg \rceil \ \Bigg)$

where $n_f$, $m$, and $N$ respectively denote the number of features, number of output classes/labels, and the total number of training samples.

\bibliographystyle{elsarticle-num}

\begin{thebibliography}{10}
\expandafter\ifx\csname url\endcsname\relax
  \def\url#1{\texttt{#1}}\fi
\expandafter\ifx\csname urlprefix\endcsname\relax\def\urlprefix{URL }\fi
\expandafter\ifx\csname href\endcsname\relax
  \def\href#1#2{#2} \def\path#1{#1}\fi

\bibitem{Bustos:2020}
O.~Bustos, A.~Pomares-Quimbaya, Stock market movement forecast: A systematic review, Expert Systems with Applications 156 (2020) 113464.

\bibitem{Kumbure:Lohrmann:2022}
M.~M. Kumbure, C.~Lohrmann, P.~Luukka, J.~Porras, Machine learning techniques and data for stock market forecasting: A literature review, Expert Systems with Applications 197 (2022) 116659.

\bibitem{Htun:Biehl:2023}
H.~H. Htun, M.~Biehl, N.~Petkov, Survey of feature selection and extraction techniques for stock market prediction, Financial Innovation 9~(1) (2023) 26.

\bibitem{Henrique_EtAl_Review_Market_prediction_snooping_2019}
B.~M. Henrique, V.~A. Sobreiro, H.~Kimura, Literature review: Machine learning techniques applied to financial market prediction, Expert Systems with Applications 124 (2019) 226--251.

\bibitem{Atsalakis:Valavanis:2009}
G.~S. Atsalakis, K.~P. Valavanis, Surveying stock market forecasting techniques – {Part II}: Soft computing methods, Expert Systems with Applications 36~(3, Part 2) (2009) 5932--5941.

\bibitem{Hussain:Knowles:2008}
A.~J. Hussain, A.~Knowles, P.~J. Lisboa, W.~El-Deredy, Financial time series prediction using polynomial pipelined neural networks, Expert Systems with Applications 35~(3) (2008) 1186--1199.

\bibitem{Lam:2004}
M.~Lam, Neural network techniques for financial performance prediction: integrating fundamental and technical analysis, Decision Support Systems 37~(4) (2004) 567--581.

\bibitem{Versace:Bhatt:2004}
M.~Versace, R.~Bhatt, O.~Hinds, M.~Shiffer, Predicting the exchange traded fund {DIA} with a combination of genetic algorithms and neural networks, Expert Systems with Applications 27~(3) (2004) 417--425.

\bibitem{Yao:1999}
X.~Yao, Evolving artificial neural networks, Proceedings of the IEEE 87~(9) (1999) 1423--1447.

\bibitem{Wang:Xu:2018}
J.~Wang, C.~Xu, X.~Yang, J.~M. Zurada, A novel pruning algorithm for smoothing feedforward neural networks based on group lasso method, IEEE Transactions on Neural Networks and Learning Systems 29~(5) (2018) 2012--2024.

\bibitem{Kohavi:John:1997}
R.~Kohavi, G.~H. John, Wrappers for feature subset selection, Artificial intelligence 97~(1) (1997) 273--324.

\bibitem{Guyon:Isabelle:2003}
I.~Guyon, A.~Elisseeff, An introduction to variable and feature selection, Journal of Machine Learning Research 3~(Mar) (2003) 1157--1182.

\bibitem{Hafiz:Swain:2018}
F.~Hafiz, A.~Swain, N.~Patel, C.~Naik, A two-dimensional ({2D}) learning framework for particle swarm based feature selection, Pattern Recognition 76 (2018) 416--433.

\bibitem{Peng:Albuquerque:2021}
Y.~Peng, P.~H.~M. Albuquerque, H.~Kimura, C.~A. P.~B. Saavedra, Feature selection and deep neural networks for stock price direction forecasting using technical analysis indicators, Machine Learning with Applications 5 (2021) 100060.

\bibitem{Asadi2012}
S.~Asadi, E.~Hadavandi, F.~Mehmanpazir, M.~M. Nakhostin, Hybridization of evolutionary levenberg–marquardt neural networks and data pre-processing for stock market prediction, Knowledge-Based Systems 35 (2012) 245--258.

\bibitem{Qiu:Song:2016}
M.~Qiu, Y.~Song, F.~Akagi, Application of artificial neural network for the prediction of stock market returns: The case of the japanese stock market, Chaos, Solitons \& Fractals 85 (2016) 1--7.

\bibitem{Zhong:Enke:2017}
X.~Zhong, D.~Enke, Forecasting daily stock market return using dimensionality reduction, Expert Systems with Applications 67 (2017) 126--139.

\bibitem{Tsai:Hsiao:2010}
C.-F. Tsai, Y.-C. Hsiao, Combining multiple feature selection methods for stock prediction: Union, intersection, and multi-intersection approaches, Decision Support Systems 50~(1) (2010) 258--269.

\bibitem{Zhang:Chen:2004}
Q.~Zhang, J.~C. Chen, P.~P. Chong, Decision consolidation: criteria weight determination using multiple preference formats, Decision Support Systems 38~(2) (2004) 247--258.

\bibitem{Parreiras:Vasconcelos:2009}
R.~Parreiras, J.~Vasconcelos, Decision making in multiobjective optimization aided by the multicriteria tournament decision method, Nonlinear Analysis: Theory, Methods \& Applications 71~(12) (2009) e191--e198.

\bibitem{Deb:Pratap:2002}
K.~Deb, A.~Pratap, S.~Agarwal, T.~Meyarivan, A fast and elitist multiobjective genetic algorithm: {NSGA-II}, IEEE Transactions on Evolutionary Computation 6~(2) (2002) 182--197.

\bibitem{Cai:Li:2015}
X.~Cai, Y.~Li, Z.~Fan, Q.~Zhang, An external archive guided multiobjective evolutionary algorithm based on decomposition for combinatorial optimization, IEEE Transactions on Evolutionary Computation 19~(4) (2015) 508--523.

\bibitem{Ishibuchi:Tsukamoto:2010}
H.~{Ishibuchi}, N.~{Tsukamoto}, Y.~{Nojima}, Diversity improvement by non-geometric binary crossover in evolutionary multiobjective optimization, IEEE Transactions on Evolutionary Computation 14~(6) (2010) 985--998.

\bibitem{Li:Zhang:2022}
G.~Li, A.~Zhang, Q.~Zhang, D.~Wu, C.~Zhan, Pearson correlation coefficient-based performance enhancement of broad learning system for stock price prediction, IEEE Transactions on Circuits and Systems II: Express Briefs 69~(5) (2022) 2413--2417.

\bibitem{kumar:Meghwani:2016}
D.~Kumar, S.~S. Meghwani, M.~Thakur, Proximal support vector machine based hybrid prediction models for trend forecasting in financial markets, Journal of Computational Science 17 (2016) 1--13.

\bibitem{Zbikowski:2015}
K.~Żbikowski, Using volume weighted support vector machines with walk forward testing and feature selection for the purpose of creating stock trading strategy, Expert Systems with Applications 42~(4) (2015) 1797--1805.

\bibitem{Oliveira:Nobre:2013}
F.~A. {de Oliveira}, C.~N. Nobre, L.~E. Zárate, Applying artificial neural networks to prediction of stock price and improvement of the directional prediction index – case study of petr4, petrobras, brazil, Expert Systems with Applications 40~(18) (2013) 7596--7606.

\bibitem{Huang2009}
C.-L. Huang, C.-Y. Tsai, A hybrid sofm-svr with a filter-based feature selection for stock market forecasting, Expert Systems with Applications 36~(2, Part 1) (2009) 1529--1539.

\bibitem{Sun:Xiao:2019}
J.~Sun, K.~Xiao, C.~Liu, W.~Zhou, H.~Xiong, Exploiting intra-day patterns for market shock prediction: A machine learning approach, Expert Systems with Applications 127 (2019) 272--281.

\bibitem{Enke:2004}
S.~Thawornwong, D.~Enke, The adaptive selection of financial and economic variables for use with artificial neural networks, Neurocomputing 56 (2004) 205--232.

\bibitem{Lee:2009}
M.-C. Lee, Using support vector machine with a hybrid feature selection method to the stock trend prediction, Expert Systems with Applications 36~(8) (2009) 10896--10904.

\bibitem{Weng:Ahmed:2017}
B.~Weng, M.~A. Ahmed, F.~M. Megahed, Stock market one-day ahead movement prediction using disparate data sources, Expert Systems with Applications 79 (2017) 153--163.

\bibitem{Inthachot:Boonjing:2016}
M.~Inthachot, V.~Boonjing, S.~Intakosum, et~al., Artificial neural network and genetic algorithm hybrid intelligence for predicting {Thai} stock price index trend, Computational intelligence and neuroscience 2016 (2016).

\bibitem{Liu:Wang:2019}
G.~Liu, X.~Wang, A new metric for individual stock trend prediction, Engineering Applications of Artificial Intelligence 82 (2019) 1--12.

\bibitem{Zhong:Enke:2017b}
X.~Zhong, D.~Enke, A comprehensive cluster and classification mining procedure for daily stock market return forecasting, Neurocomputing 267 (2017) 152--168.

\bibitem{Persio:Honchar:2016}
L.~Di~Persio, O.~Honchar, Artificial neural networks architectures for stock price prediction: Comparisons and applications, International journal of circuits, systems and signal processing 10~(2016) (2016) 403--413.

\bibitem{UlHaq2021}
A.~U. Haq, A.~Zeb, Z.~Lei, D.~Zhang, Forecasting daily stock trend using multi-filter feature selection and deep learning, Expert Systems with Applications 168 (2021) 114444.

\bibitem{Alsubaie:Hindi:2019}
Y.~Alsubaie, K.~El~Hindi, H.~Alsalman, Cost-sensitive prediction of stock price direction: Selection of technical indicators, IEEE Access 7 (2019) 146876--146892.

\bibitem{Sezer:Gudelek:2020}
O.~B. Sezer, M.~U. Gudelek, A.~M. Ozbayoglu, Financial time series forecasting with deep learning: A systematic literature review: 2005--2019, Applied soft computing 90 (2020) 106181.

\bibitem{Rundo:Trenta:2019}
F.~Rundo, F.~Trenta, A.~L. di~Stallo, S.~Battiato, Machine learning for quantitative finance applications: A survey, Applied Sciences 9~(24) (2019) 5574.

\bibitem{he2015deep}
K.~He, X.~Zhang, S.~Ren, J.~Sun, Deep residual learning for image recognition, in: Proceedings of the IEEE conference on computer vision and pattern recognition, 2016, pp. 770--778.

\bibitem{Fama1965}
E.~F. Fama, Random walks in stock market prices, Financial Analysts Journal 21~(5) (1965) 55--59.

\bibitem{Stathakis:2009}
D.~Stathakis, How many hidden layers and nodes?, International Journal of Remote Sensing 30~(8) (2009) 2133--2147.

\bibitem{Hafiz:Broekaert:2021}
F.~Hafiz, J.~Broekaert, D.~{La Torre}, A.~Swain, A multi-criteria approach to evolve sparse neural architectures for stock market forecasting, Annals of Operations Research (Accepted), Available online: arXiv eprint:2111.08060 (2021).

\bibitem{Kim2003}
K.~Kim, Financial time series forecasting using support vector machines, Neurocomputing 55~(1) (2003) 307--319.

\bibitem{Cao:Tay:2003}
L.-J. Cao, F.~E.~H. Tay, Support vector machine with adaptive parameters in financial time series forecasting, IEEE Transactions on {N}eural {N}etworks 14~(6) (2003) 1506--1518.

\bibitem{Olson:Mossman:2003}
D.~Olson, C.~Mossman, Neural network forecasts of canadian stock returns using accounting ratios, International Journal of Forecasting 19~(3) (2003) 453--465.

\bibitem{Hu:tang:2018}
H.~Hu, L.~Tang, S.~Zhang, H.~Wang, Predicting the direction of stock markets using optimized neural networks with google trends, Neurocomputing 285 (2018) 188--195.

\bibitem{Chang2012}
P.-C. Chang, D.~di~Wang, C.~le~Zhou, A novel model by evolving partially connected neural network for stock price trend forecasting, Expert Systems with Applications 39~(1) (2012) 611--620.

\bibitem{Lei2018}
L.~Lei, Wavelet neural network prediction method of stock price trend based on rough set attribute reduction, Applied Soft Computing 62 (2018) 923--932.

\bibitem{Farahani:2021}
M.~Shahvaroughi~Farahani, S.~H. Razavi~Hajiagha, Forecasting stock price using integrated artificial neural network and metaheuristic algorithms compared to time series models, Soft computing 25~(13) (2021) 8483--8513.

\bibitem{Gocken:2016}
M.~Göçken, M.~Özçalıcı, A.~Boru, A.~T. Dosdoğru, Integrating metaheuristics and artificial neural networks for improved stock price prediction, Expert Systems with Applications 44 (2016) 320--331.

\bibitem{Kim:Shin:2007}
H.~jung Kim, K.~shik Shin, A hybrid approach based on neural networks and genetic algorithms for detecting temporal patterns in stock markets, Applied Soft Computing 7~(2) (2007) 569--576.

\bibitem{Shynkevich:McGinnity:2017}
Y.~Shynkevich, T.~McGinnity, S.~Coleman, A.~Belatreche, Y.~Li, Forecasting price movements using technical indicators: Investigating the impact of varying input window length, Neurocomputing 264 (2017) 71--88.

\bibitem{Gao_EtAl_Covid19_volatility_US_China_2022}
X.~Gao, Y.~Ren, M.~Umar, To what extent does {COVID}-19 drive stock market volatility? a comparison between the {U.S.} and china, Economic Research 35~(1) (2022) 1686--1706.

\bibitem{Rahman_EtAL_COVID19_volatility_2022}
M.~M. Rahman, C.~Guotai, A.~D. Gupta, M.~Hossain, M.~Z. Abedin, Impact of early {COVID}-19 pandemic on the {US} and european stock markets and volatility forecasting, Economic Research 35~(1) (2022) 3591--3608.

\bibitem{Li_EtAl_Covid19_fear_volatility_2022}
W.~Li, F.~Chien, H.~W. Kamran, T.~M. Aldeehani, M.~Sadiq, V.~C. Nguyen, F.~Taghizadeh-Hesary, The nexus between {COVID}-19 fear and stock market volatility, Economic Research 35~(1) (2022) 1765--1785.

\bibitem{BuszkoEtAl2021}
M.~Buszko, W.~Orzeszko, M.~Stawarz, {COVID}-19 pandemic and stability of stock market—a sectoral approach, PLOS ONE 16~(5) (2021) 1--26.

\bibitem{ChandraEtAl2021}
R.~Chandra, Y.~He, Bayesian neural networks for stock price forecasting before and during {COVID}-19 pandemic, PLOS ONE 16~(7) (2021) 1--32.

\bibitem{Sullivan_EtAl_data_snooping_technical_trading_1999}
R.~Sullivan, A.~Timmermann, H.~White, Data-snooping, technical trading rule performance, and the bootstrap, The Journal of Finance 54~(5) (1999) 1647--1691.

\bibitem{Grandini:Bagli:2020}
M.~Grandini, E.~Bagli, G.~Visani, Metrics for multi-class classification: an overview (2020).
\newblock \href {https://doi.org/10.48550/ARXIV.2008.05756} {\path{doi:10.48550/ARXIV.2008.05756}}.

\bibitem{Hagg:Mensing:2017}
A.~Hagg, M.~Mensing, A.~Asteroth, Evolving parsimonious networks by mixing activation functions, in: Proceedings of the Genetic and Evolutionary Computation Conference, 2017, pp. 425--432.

\bibitem{Hafiz:Swain:MOEA:2020}
F.~Hafiz, A.~Swain, E.~Mendes, Multi-objective evolutionary framework for non-linear system identification: A comprehensive investigation, Neurocomputing 386 (2020) 257 -- 280.

\bibitem{Moller:1993}
M.~F. M{\o}ller, A scaled conjugate gradient algorithm for fast supervised learning, Neural Networks 6~(4) (1993) 525--533.

\bibitem{Peng:Long:2005}
H.~Peng, F.~Long, C.~Ding, Feature selection based on mutual information criteria of max-dependency, max-relevance, and min-redundancy, IEEE {T}ransactions on {P}attern {A}nalysis and {M}achine {I}ntelligence 27~(8) (2005) 1226--1238.

\bibitem{Hall:1999}
M.~A. Hall, L.~A. Smith, Feature selection for machine learning: comparing a correlation-based filter approach to the wrapper., in: FLAIRS conference, Vol. 1999, 1999, pp. 235--239.

\bibitem{Garcia:Salvador:2010}
S.~Garc{\'\i}a, A.~Fern{\'a}ndez, J.~Luengo, F.~Herrera, Advanced nonparametric tests for multiple comparisons in the design of experiments in computational intelligence and data mining: Experimental analysis of power, Information Sciences 180~(10) (2010) 2044--2064.

\end{thebibliography}

\end{document}